\input harvmac
\let\includefigures=\iftrue
\let\useblackboard=\iftrue
\newfam\black

\includefigures
\message{If you do not have epsf.tex (to include figures),}
\message{change the option at the top of the tex file.}
\input epsf
\def\figin{\epsfcheck\figin}\def\figins{\epsfcheck\figins}
\def\epsfcheck{\ifx\epsfbox\UnDeFiNeD
\message{(NO epsf.tex, FIGURES WILL BE IGNORED)}
\gdef\figin##1{\vskip2in}\gdef\figins##1{\hskip.5in}
\else\message{(FIGURES WILL BE INCLUDED)}%
\gdef\figin##1{##1}\gdef\figins##1{##1}\fi}
\def\DefWarn#1{}
\def\figinsert{\goodbreak\midinsert}
\def\ifig#1#2#3{\DefWarn#1\xdef#1{fig.~\the\figno}
\writedef{#1\leftbracket fig.\noexpand~\the\figno}%
\figinsert\figin{\centerline{#3}}\medskip\centerline{\vbox{
\baselineskip12pt\advance\hsize by -1truein
\noindent\footnotefont{\bf Fig.~\the\figno:} #2}}
\endinsert\global\advance\figno by1}
\else
\def\ifig#1#2#3{\xdef#1{fig.~\the\figno}
\writedef{#1\leftbracket fig.\noexpand~\the\figno}%
\global\advance\figno by1} \fi
%

\def\id{{1 \kern-.28em {\rm l}}}

\def\Z{{\bf Z}}

\def\K3{{\bf K3}}
\def\journal#1&#2(#3){\unskip, \sl #1\ \bf #2 \rm(19#3) }
\def\andjournal#1&#2(#3){\sl #1~\bf #2 \rm (19#3) }

\def\bar{\overline}
\def\hat{\widehat}
\def\ie{{\it i.e.}}

\def\etc{{\it etc}}

\def\tilde{\widetilde}

\def\frac#1#2{{#1\over#2}}

\def\half{\frac12}

\def\d{\partial}

\def\inbar{\,\vrule height1.5ex width.4pt depth0pt}
\def\IC{\relax\hbox{$\inbar\kern-.3em{\rm C}$}}
\def\Ic{\relax\hbox{$\inbar\kern-.3em{\rm c}$}}
\def\IR{\relax{\rm I\kern-.18em R}}
\def\IP{\relax{\rm I\kern-.18em P}}
\def\Z{{\bf Z}}

%
%

%
\catcode`\@=11
\def\slash#1{\mathord{\mathpalette\c@ncel{#1}}}
\overfullrule=0pt

\def\FF{{\cal F}}

\def\JJ{{\cal J}}
\def\KK{{\cal K}}
\def\LL{{\cal L}}

\def\OO{{\cal O}}

\def\underrel#1\over#2{\mathrel{\mathop{\kern\z@#1}\limits_{#2}}}

\catcode`\@=12


%

\def\det{{\rm det}}

\def \sgn{{\rm sgn}}
\def\det{{\rm det}}


\def\xbar{{\bar x}}
\def\zbar{{\bar z}}

\def\abar{{\bar a}}
\def\bbar{{\bar b}}
\def\cbar{{\bar c}}
\def\dbar{{\bar d}}
\def\ebar{{\bar e}}
\def\fbar{{\bar f}}

\def\hbar{{\bar h}}

\def\vbar{{\bar v}}
\def\wbar{{\overline w}}

\def\delbar{{\bar \del}}


\lref\SeibergZK{
  N.~Seiberg,
  Phys.\ Lett.\  B {\bf 408}, 98 (1997)
  [arXiv:hep-th/9705221].
}

\lref\AharonyUB{
  O.~Aharony, M.~Berkooz, D.~Kutasov and N.~Seiberg,
  JHEP {\bf 9810}, 004 (1998)
  [arXiv:hep-th/9808149].
}

\lref\HananyHQ{
  A.~Hanany and I.~R.~Klebanov,
  Nucl.\ Phys.\  B {\bf 482}, 105 (1996)
  [arXiv:hep-th/9606136].
}

\lref\CallanAT{
  C.~G.~Callan, J.~A.~Harvey and A.~Strominger,
  ``Supersymmetric string solitons,''
  arXiv:hep-th/9112030.
}

\lref\TongRQ{
  D.~Tong,
  JHEP {\bf 0207}, 013 (2002)
  [arXiv:hep-th/0204186].
}

\lref\HarveyAB{
  J.~A.~Harvey and S.~Jensen,
  JHEP {\bf 0510}, 028 (2005)
  [arXiv:hep-th/0507204].
}

\lref\OkuyamaGX{
  K.~Okuyama,
  JHEP {\bf 0508}, 089 (2005)
  [arXiv:hep-th/0508097].
}

\lref\WittenXU{
  E.~Witten,
  JHEP {\bf 0906}, 067 (2009)
  [arXiv:0902.0948 [hep-th]].
}

\lref\BuscherSK{
  T.~H.~Buscher,
  Phys.\ Lett.\  B {\bf 194}, 59 (1987).
}
\lref\BergshoeffAS{
  E.~Bergshoeff, C.~M.~Hull and T.~Ortin,
  Nucl.\ Phys.\  B {\bf 451}, 547 (1995)
  [arXiv:hep-th/9504081].
}

\lref\BershadskySP{
  M.~Bershadsky, C.~Vafa and V.~Sadov,
  Nucl.\ Phys.\  B {\bf 463}, 398 (1996)
  [arXiv:hep-th/9510225].
}

\lref\DasguptaSU{
  K.~Dasgupta and S.~Mukhi,
  Nucl.\ Phys.\  B {\bf 551}, 204 (1999)
  [arXiv:hep-th/9811139].
}

\lref\UrangaVF{
  A.~M.~Uranga,
  JHEP {\bf 9901}, 022 (1999)
  [arXiv:hep-th/9811004].
}

\lref\HananyIT{
  A.~Hanany and A.~M.~Uranga,
  JHEP {\bf 9805}, 013 (1998)
  [arXiv:hep-th/9805139].
}

\lref\AndreasHH{
  B.~Andreas, G.~Curio and D.~Lust,
  JHEP {\bf 9810}, 022 (1998)
  [arXiv:hep-th/9807008].
}

\lref\AganagicFE{
  M.~Aganagic, A.~Karch, D.~Lust and A.~Miemiec,
  Nucl.\ Phys.\  B {\bf 569}, 277 (2000)
  [arXiv:hep-th/9903093].
}

\lref\BeckerQH{
  M.~Becker, K.~Dasgupta, A.~Knauf and R.~Tatar,
  Nucl.\ Phys.\  B {\bf 702}, 207 (2004)
  [arXiv:hep-th/0403288];
  M.~Becker, K.~Dasgupta, S.~H.~Katz, A.~Knauf and R.~Tatar,
  Nucl.\ Phys.\  B {\bf 738}, 124 (2006)
  [arXiv:hep-th/0511099].
}

\lref\HananyIE{
  A.~Hanany and E.~Witten,
  Nucl.\ Phys.\  B {\bf 492}, 152 (1997)
  [arXiv:hep-th/9611230].
}

\lref\ElitzurFH{
  S.~Elitzur, A.~Giveon and D.~Kutasov,
  Phys.\ Lett.\  B {\bf 400}, 269 (1997)
  [arXiv:hep-th/9702014].
}

\lref\HananyTB{
  A.~Hanany and A.~Zaffaroni,
  JHEP {\bf 9805}, 001 (1998)
  [arXiv:hep-th/9801134].
}

\lref\CandelasJS{
  P.~Candelas and X.~C.~de la Ossa,
  Nucl.\ Phys.\  B {\bf 342}, 246 (1990).
}

\lref\TseytlinBH{
  A.~A.~Tseytlin,
  Nucl.\ Phys.\  B {\bf 475}, 149 (1996)
  [arXiv:hep-th/9604035].
}

\lref\TseytlinCS{
  A.~A.~Tseytlin,
  Class.\ Quant.\ Grav.\  {\bf 14}, 2085 (1997)
  [arXiv:hep-th/9702163].
}

\lref\PolchinskiRR{
  J.~Polchinski,
{\it  Cambridge, UK: Univ. Pr. (1998) 531 p}
}

\lref\CherkisIR{
  S.~A.~Cherkis and A.~Hashimoto,
  JHEP {\bf 0211}, 036 (2002)
  [arXiv:hep-th/0210105].
}

\lref\ArapogluAH{
  S.~Arapoglu, N.~S.~Deger and A.~Kaya,
  Phys.\ Lett.\  B {\bf 578}, 203 (2004)
  [arXiv:hep-th/0306040].
}

\lref\GauntlettPB{
  J.~P.~Gauntlett, D.~A.~Kastor and J.~H.~Traschen,
  Nucl.\ Phys.\  B {\bf 478}, 544 (1996)
  [arXiv:hep-th/9604179].
}

\lref\KhuriII{
  R.~R.~Khuri,
  Phys.\ Rev.\  D {\bf 48}, 2947 (1993)
  [arXiv:hep-th/9305143].
}

\lref\LinNB{
  H.~Lin, O.~Lunin and J.~M.~Maldacena,
  JHEP {\bf 0410}, 025 (2004)
  [arXiv:hep-th/0409174].
}

\lref\DHokerXY{
E.~D'Hoker, J.~Estes and M.~Gutperle,
  JHEP {\bf 0706}, 021 (2007)
  [arXiv:0705.0022 [hep-th]];
  E.~D'Hoker, J.~Estes and M.~Gutperle,
  JHEP {\bf 0706}, 022 (2007)
  [arXiv:0705.0024 [hep-th]].
}

\lref\DHokerWC{
  E.~D'Hoker, J.~Estes, M.~Gutperle and D.~Krym,
  JHEP {\bf 0808}, 028 (2008)
  [arXiv:0806.0605 [hep-th]];
  E.~D'Hoker, J.~Estes, M.~Gutperle and D.~Krym,
  JHEP {\bf 0812}, 044 (2008)
  [arXiv:0810.4647 [hep-th]];
  E.~D'Hoker, J.~Estes, M.~Gutperle and D.~Krym,
  JHEP {\bf 0909}, 067 (2009)
  [arXiv:0906.0596 [hep-th]].
}

\lref\ItzhakiUZ{
  N.~Itzhaki, A.~A.~Tseytlin and S.~Yankielowicz,
  Phys.\ Lett.\  B {\bf 432}, 298 (1998)
  [arXiv:hep-th/9803103].
}

\lref\HashimotoUG{
  A.~Hashimoto,
  JHEP {\bf 9901}, 018 (1999)
  [arXiv:hep-th/9812159].
}

\lref\FayyazuddinEM{
  A.~Fayyazuddin and D.~J.~Smith,
  JHEP {\bf 0010}, 023 (2000)
  [arXiv:hep-th/0006060].
}

\lref\LuninTF{
  O.~Lunin,
  JHEP {\bf 0809}, 028 (2008)
  [arXiv:0802.0735 [hep-th]].
}

\lref\FayyazuddinZU{
  A.~Fayyazuddin and D.~J.~Smith,
  JHEP {\bf 9904}, 030 (1999)
  [arXiv:hep-th/9902210].
}

\lref\GomberoffPS{
  A.~Gomberoff, D.~Kastor, D.~Marolf and J.~H.~Traschen,
  Phys.\ Rev.\  D {\bf 61}, 024012 (2000)
  [arXiv:hep-th/9905094].
}

\lref\PandoZayasSQ{
  L.~A.~Pando Zayas and A.~A.~Tseytlin,
  JHEP {\bf 0011}, 028 (2000)
  [arXiv:hep-th/0010088].
}

\lref\GregoryTE{
  R.~Gregory, J.~A.~Harvey and G.~W.~Moore,
  Adv.\ Theor.\ Math.\ Phys.\  {\bf 1}, 283 (1997)
  [arXiv:hep-th/9708086].
}

\lref\LindstromRT{
  U.~Lindstrom and M.~Rocek,
  Nucl.\ Phys.\  B {\bf 222}, 285 (1983).
}

\lref\HitchinEA{
  N.~J.~Hitchin, A.~Karlhede, U.~Lindstrom and M.~Rocek,
  Commun.\ Math.\ Phys.\  {\bf 108}, 535 (1987).
}

\lref\RocekPS{
  M.~Rocek and E.~P.~Verlinde,
  Nucl.\ Phys.\  B {\bf 373}, 630 (1992)
  [arXiv:hep-th/9110053].
}

\lref\KiritsisWD{
  E.~Kiritsis, C.~Kounnas and D.~Lust,
  arXiv:hep-th/9312143.
}

\lref\BalasubramanianDV{
  A.~K.~Balasubramanian, S.~Govindarajan and C.~N.~Gowdigere,
  Class.\ Quant.\ Grav.\  {\bf 24}, 6393 (2007)
  [arXiv:0707.4306 [hep-th]].
}

\lref\Abreu{
  M.~Abreu,
  Toric Varieties in Algebraic Geometry and Physics, AMS,
  [arXiv:math/0004122].
}

\lref\ChenDU{
  B.~Chen {\it et al.},
  JHEP {\bf 0710}, 003 (2007)
  [arXiv:0704.2233 [hep-th]].
}

\lref\MR{
  J.~McOrist and A.~B.~Royston,
  in preparation.
}

\lref\MorrisonCS{
  D.~R.~Morrison and M.~R.~Plesser,
  Adv.\ Theor.\ Math.\ Phys.\  {\bf 3}, 1 (1999)
  [arXiv:hep-th/9810201].
}

\lref\BuicanIS{
  M.~Buican, D.~Malyshev and H.~Verlinde,
  JHEP {\bf 0806}, 108 (2008)
  [arXiv:0710.5519 [hep-th]].
}

\lref\GiveonUR{
  A.~Giveon, D.~Kutasov, J.~McOrist and A.~B.~Royston,
  Nucl.\ Phys.\  B {\bf 822}, 106 (2009)
  [arXiv:0904.0459 [hep-th]].
}

\lref\KutasovKB{
  D.~Kutasov, O.~Lunin, J.~McOrist and A.~B.~Royston,
  Nucl.\ Phys.\  B {\bf 833}, 64 (2010)
  [arXiv:0909.3319 [hep-th]].
}

\lref\MaldacenaMW{
  J.~Maldacena and D.~Martelli,
  JHEP {\bf 1001}, 104 (2010)
  [arXiv:0906.0591 [hep-th]].
}

\lref\BanksRJ{
  T.~Banks, M.~Dine, H.~Dykstra and W.~Fischler,
  Phys.\ Lett.\  B {\bf 212}, 45 (1988).
}

\lref\BakasBA{
  I.~Bakas,
  Phys.\ Lett.\  B {\bf 343}, 103 (1995)
  [arXiv:hep-th/9410104].
}

\lref\AlvarezZR{
  E.~Alvarez, L.~Alvarez-Gaume and I.~Bakas,
  Nucl.\ Phys.\  B {\bf 457}, 3 (1995)
  [arXiv:hep-th/9507112].
}

\lref\AlvarezAI{
  E.~Alvarez, L.~Alvarez-Gaume and I.~Bakas,
  Nucl.\ Phys.\ Proc.\ Suppl.\  {\bf 46}, 16 (1996)
  [arXiv:hep-th/9510028].
}

\rightline{DAMTP-2011-4}
\rightline{RUNHETC-2010-27}

\Title{}
{\vbox{\centerline{Relating Conifold Geometries to $NS5$-branes}
\bigskip
\centerline{}
}}
\bigskip

\centerline{\it Jock McOrist$^{1}$ and Andrew B. Royston$^{2}$}
\bigskip
\smallskip
\centerline{${}^{1}$DAMTP, Centre for Mathematical Sciences}
\centerline{ Wilberforce Road, Cambridge, CB3 OWA, UK}
\smallskip
\centerline{${}^{2}$NHETC and Department of Physics and Astronomy} \centerline{Rutgers University, Piscataway, NJ 08855, USA }
\smallskip

\smallskip

\vglue .3cm

\bigskip

\let\includefigures=\iftrue
\bigskip
\noindent
We construct the first known example of a near horizon supergravity solution for a pair of $NS5$-branes, intersecting on $\IR^{1,3}$ and localised in all directions except a single transverse circle. We do this by establishing an explicit map between the conifold metric and the near horizon geometry of two intersecting $NS5$-branes, clarifying and correcting a number of open issues in the literature en route.  Our technique is general in nature and may be applied to a whole class of 1/4-BPS five-brane webs and their geometric duals.  These 1/4-BPS solutions may have an interesting holographic interpretation in terms of little string theory.

\bigskip

\Date{}

\newsec{Introduction}

Examples of 1/4-BPS supergravity solutions have served as an important tool in many areas of string theory. Most recently, these solutions have been crucial in the development of gauge/gravity duality, in which the near horizon limit of intersecting $D$-branes serves as a holographic dual to certain supersymmetric field theories. Within this paradigm most known 1/4-BPS solutions involve $D$-branes only; little is known about their $NS5$-brane cousins, even though their geometries may shed light on the definition of little string theory \refs{\SeibergZK,\HananyHQ,\AharonyUB}. In this paper we construct the 1/4-BPS near horizon geometry of a pair of intersecting $NS5$-branes, localised in all directions except a single transverse circular direction\foot{At the level of supergravity this is the best one can do. Indeed, the T-dual of the Taub-NUT metric is the CHS geometry smeared in the circle direction \refs{\BanksRJ,\CallanAT}. The localisation along that circle, which is related to modes of the $B$-field on the Taub-NUT side, only occurs once one includes worldsheet instanton effects \refs{\TongRQ,\HarveyAB,\OkuyamaGX}.  Alternatively the relationship can be deduced by considering probe strings \GregoryTE\ or $D$-branes \WittenXU, but either way, it is beyond the applicability of the Buscher rules \refs{\BuscherSK,\BergshoeffAS} for T-duality at the supergravity level.}. Along the way we clarify the T-duality relation between the conifold and intersecting $NS5$-branes. For example, we show the correct choice of U(1) isometry along which one T-dualises is subtle, and the duality map also requires a Legendre transformation on the metric.  Furthermore, the T-dual of the conifold is not the complete 1/4-BPS supergravity solution for intersecting $NS5$-branes. Rather, one arrives at a near horizon limit.

There is a great body of literature on duality relations between string theory in conifold geometries and string theory in the presence of intersecting $NS5$-branes, starting with the work of \BershadskySP.  The conifold possesses three $U(1)$ isometry directions along which one may T-dualise.  Starting with (say) type IIB string theory in the conifold background, a single T-duality along any of these directions takes one to type IIA in the presence of two $NS5$-branes intersecting in $1+3$ dimensions \refs{\DasguptaSU,\UrangaVF}, two T-dualities takes one to IIB string theory in the presence of intersecting $NS5$-branes \refs{\BershadskySP,\HananyIT,\AndreasHH}, and T-duality along all three isometry directions is related to mirror symmetry \refs{\AganagicFE,\BeckerQH}.

In most cases, these dualities were discovered and investigated with the aid of $D$-branes.  $D$-branes that wrap shrinking cycles represent BPS excitations that are becoming massless.  In T-dual pictures, these branes map to BPS configurations of $D$-branes stretched between $NS5$-branes of the sort considered in \refs{\HananyIE,\ElitzurFH,\HananyTB}, that may also possess massless excitations as the moduli are varied.  Matching the spectrum of massless modes provides a nontrivial consistency check of the duality relation.

However, to the best of our knowledge, an explicit demonstration of the T-duality relation between the conifold and two intersecting $NS5$-branes at the level of supergravity has never been fully realised.  The first attempt at addressing this issue was made by \DasguptaSU.  They started with a delocalised version of the IIA brane configuration in which the $NS5$-branes are smeared in all transverse directions except one.  The supergravity solution for the smeared configuration can be readily constructed from the prescription of \refs{\TseytlinBH,\TseytlinCS}.  By T-dualising along one of the directions in which the branes are smeared, \DasguptaSU\ arrived at a space with a similar fibre structure to the conifold, but where the $S^1$ fibration is over $\IR^2 \times \IR^2$ instead of $S^2 \times S^2$.  We show that this discrepancy is due to the smearing of the $NS5$-branes in the relatively transverse directions.  Furthermore, the embedding of the $U(1)$ fibres, as well as the other conifold coordinates, into the brane configuration proposed in \DasguptaSU\ requires some modification.

Explicit supergravity solutions for fully localised brane intersections are rare. The nature of the intersection may be classified by the boundary conditions on open strings following the language of \PolchinskiRR\foot{Strictly speaking this labeling describes boundary conditions on open strings ending on $D$-branes; in a slight abuse of notation we will use it to classify possible membrane intersections in supergravity.}. For a pair of branes, there are three possible transversal directions: Dirichlet-Dirchlet (DD), Dirichlet-Neumann (DN), and Neumann-Dirichlet (ND). The known intersecting brane supergravity solutions are when one brane is fully embedded in the worldvolume of the other, or when the intersection is completely transversal. The first case corresponds to having ND and DD directions but no DN directions; examples include \refs{\CherkisIR,\ArapogluAH}.  The second case corresponds to DN and ND directions but no DD directions;  examples include \refs{\TseytlinCS,\GauntlettPB,\KhuriII}.  These solutions solve the equations of motion with delta-function source terms for the branes.  There are also 1/2-BPS solutions that solve the equations of motion without sources, and represent near horizon limits of intersecting brane systems \refs{\LinNB,\DHokerXY,\DHokerWC}.  The construction of these solutions is facilitated by the (super-) symmetry enhancement that typically accompanies the near horizon limit\foot{There are also interesting ``hybrid'' solutions where a near horizon limit is taken for one set of branes, while the other set remains $\delta$-function localised \ItzhakiUZ, including a couple interesting examples with DN, ND and DD directions \refs{\HashimotoUG,\FayyazuddinEM}.}.  As far as we are aware, there are no BPS supergravity solutions with localised sources for both branes, that have all three types of transversal directions: DD, DN and ND.

We will construct a supergravity solution for two intersecting $NS5$-branes smeared on a single transverse circle; this smearing is necessary for the application of Buscher's T-duality rules \refs{\BuscherSK,\BergshoeffAS}.  The branes will be localised in all remaining directions, and this is the first explicit solution with nontrivial dependence on DN, ND, and DD directions, solving the equations of motion with source terms for each brane.  It is a near horizon solution, but only in the sense that the geometry is not asymptotically flat.  There is no symmetry enhancement and in particular, the isometry group does not contain a conformal factor.  This is consistent with the $NS5$-branes serving as a holographic description of little string theory, which has a scale $m_s$.

An outline of the remainder of our paper is as follows. In section 2 we U-dualise the results of \LuninTF\ for webs of $D5$-branes to write down an ansatz for the background corresponding to stacks of intersecting $NS5$-branes.  This confirms earlier results of \FayyazuddinZU\ and provides a consistency check on our starting point\foot{In the result of \FayyazuddinZU\ it was assumed that there exists coordinates in which the projections imposed on the Killing spinor are identical to those satisfied by corresponding probe branes in flat space. The same result was obtained in \LuninTF, and the existence of such coordinates proven, by using only an ansatz for the supergravity solution with the appropriate isometry and then solving the supersymmetry constraints and Bianchi identities.}.  The entire geometry is determined up to a real function $\KK$ that satisfies a nonlinear PDE of Monge-Ampere type whose singularity structure determines the number of branes in each stack as well as their location in the ten-dimensional target space.  We then smear this ansatz on a transverse circle and T-dualise to a pure metric geometry, determined up to the same function $\KK$.  For a given solution $\KK$, this metric should describe the conifold-like geometry that is T-dual to the corresponding brane web.

In section 3, we construct the explicit T-duality map between the conifold and a pair of intersecting $NS5$-branes.  The construction involves determining a solution to the Monge-Ampere equation with the appropriate singulariy structure for the brane pair, and thus we also obtain the supergravity solution describing the intersecting branes.  To the best of our knowledge this is the first time an explicit map (at the level of supergravity) showing the relation between the conifold and an intersecting $NS5$-brane solution localised in all but one direction has been constructed. The map illustrates a number of interesting points. For example, it shows the correct $U(1)\subset U(1)^3$ isometry of the conifold to T-dualise on. Related to this, the K\"ahler coordinate system of the conifold is Legendre transformed into a hybrid, symplectic-K\"ahler coordinate system that naturally describes the intersecting brane picture, and the potential $\KK$ is the Legendre transform of the K\"ahler potential.  We also show that this result is general: the hybrid coordinate system and metric that naturally describe the geometry T-dual to a generic $NS5$-brane web are related locally to a Ricci flat K\"ahler metric via Legendre transform.  This hybrid coordinate system is natural from the viewpoint of supersymmetry, isometries, and the equations of motion, and is none other than the one employed by \LuninTF\ in the description of brane webs.

In section 4 we conclude with some comments and directions for future work.  Appendices A and B contain some technical details, while in Appendix C we construct $\KK$ for the case of parallel $NS5$-branes and Taub-NUT.

\newsec{Metrics for intersecting $NS5$-branes and the geometric duals}

In this section we first construct a supergravity ansatz that describes stacks of intersecting $NS5$-branes (in type IIA or IIB).  In section 2.2 we smear the $NS5$-brane configuration on a transverse circle and T-dualise, finding a dual geometry that is pure metric.

\subsec{From $D5$-brane webs to intersecting $NS5$-branes}

The starting point is two stacks of $D5$-branes intersecting on $\IR^{1,3}$.  Let the first stack be extended in directions $(x^0,x^1,x^2,x^3,x^4,x^5)$ and the second stack in directions $(x^0,x^1,x^2,x^3,x^8,x^9)$.  We will denote the common spatial directions by ${\bf x} = (x^1,x^2,x^3)$, and the overall transverse directions by ${\bf y} = (x^6,x^7)$.  The relatively transverse coordinates may be combined into a complex coordinate system $z^1 = x^4 + i x^5$, $z^2 = x^8 + i x^9$, denoted collectively by $z^a$. Although for now the picture of orthogonally intersecting stacks is convenient, as we will see, the following geometry is capable of describing more general profiles---or ``webs''---in the space spanned by $(z^1,z^2)$.  After an impressive tour-de-force analysis of the supersymmetry constraints and Bianchi identities, the IIB geometry for this system was shown \LuninTF\ to be of the form
\eqn\Dfive{ \eqalign{ & (ds')^2 = e^{3 A/2} \left[ - dt^2 + d{\bf x}_{3}^2 + 2 \KK_{a \bbar} dz d\zbar^\bbar \ \right] + e^{-3A/2} d{\bf y}_{2}^2~, \cr
& e^{\varphi'-\varphi_0'} = e^{3A/2}~, \cr
& F_7' = i dt \wedge d( e^{3A} \KK_{a \bbar} dz^a \wedge d\zbar^{\bbar} ) \wedge d^3 {\bf x}~,}}
with all other fields vanishing. The primes are to distinguish \Dfive\ from the main geometry of interest of which we will be obtaining shortly. The warp factor and $z$-$\zbar$ metric are given in terms of a real function $\KK = \KK(z^a,\zbar^\abar, {\bf y})$ according to
\eqn\gAK{ \eqalign{ & \KK_{a\bbar} = \del_a \delbar_\bbar \KK~, \cr
& \frac{1}{4} e^{-3 A} =  \del_1 \delbar_{\bar 1} \KK \del_2\delbar_{\bar 2} \KK - \del_1\delbar_{\bar 2} \KK \del_2 \delbar_{\bar 1} \KK \equiv \det{(\del \delbar \KK)}~.}}
$F_7'$ is the Hodge dual of the R-R three-form flux, $F_7' = \star' F_3'$, $F_3' = d C_2'$.  We have factored out the asymptotic string coupling $g_s' = e^{\varphi_0'}$, so that the warp factor $e^{3A/2} \to 1$ far from the brane locus.  The equation of motion for the flux provides a second equation relating the warp factor and $\KK$:
\eqn\fluxeom{ \Delta_{\bf y} \KK + 2 e^{-3 A} = 0~,}
where $\Delta_{\bf y}$ is the flat-space Laplacian on $\IR_{\bf y}^{2}$.  This can be combined with \gAK\ to obtain an equation for $\KK$ alone:
\eqn\Monge{ \Delta_{\bf y} \KK + 8 \det{( \del \delbar \KK)} = 0~.}
With this result all remaining equations of motion are satisfied.

Actually, the equation of motion \fluxeom, and therefore \Monge, hold away from the sources and must be modified at the locations of the sources.  By considering a near-brane analysis of \gAK, \LuninTF\ demonstrated the existence of a unique set of coordinates $\lambda,\eta$, related holomorphically to $z^a$, $\lambda = \lambda(z^a), \eta = \eta(z^a)$, such that the brane locus may be described by $\lambda = \lambda_0$, ${\bf y} = {\bf y}_0$, and in the limit one approaches the source,
\eqn\nearbrane{\eqalign{ & \KK \rightarrow \frac{1}{2} \eta {\bar\eta} + \KK_2(\lambda,{\bar\lambda},{\bf y})~, \qquad {\rm with} \cr
& \Delta_{\bf y} \KK_2 + 4\del_\lambda \delbar_{\bar\lambda} \KK_2  = - \frac{Q_0}{2\pi} \delta^{(2)}({\bf y} - {\bf y}_0) \log{| \lambda - \lambda_0|^2}~,}}
where corrections are suppressed by powers of $|\lambda - \lambda_0|^2$.  With the aid of such coordinates, it is straightforward to write down generalisations of \fluxeom, \Monge\ that include source terms \refs{\LuninTF}.  Equation \fluxeom\ becomes
\eqn\fluxsource{ \del_a \delbar_\bbar  \left( \Delta_{\bf y} \KK + 2e^{-3A} \right) = - Q_0 \del_a \lambda \delbar_\bbar {\bar\lambda}  \ \delta^{(2)}({\bf y} - {\bf y}_0) \delta^{(2)}(\lambda - \lambda_0)~,}
which, together with \gAK, implies
\eqn\Mongesource{  \del_a \delbar_\bbar \left( \Delta_{\bf y} \KK + 8 \det{( \del \delbar \KK)}  \right) = - Q_0 \del_a \lambda \delbar_\bbar {\bar\lambda} \ \delta^{(2)}({\bf y} - {\bf y}_0) \delta^{(2)}(\lambda - \lambda_0)~.}
The solution of \Mongesource, plugged into \Dfive\ provides the supergravity background corresponding to a web of $D5$-branes following a holomorphic profile $\lambda(z) = \lambda_0$, situated at ${\bf y} = {\bf y}_0$.  Multiple webs at different positions, $(\lambda_i,{\bf y}_i)$, are mutually BPS, and equations \fluxsource, \Mongesource\ may be generalised to this situation by superposing the sources\foot{The existence of solutions to \Mongesource\ is an interesting question. \LuninTF, correcting an error in \GomberoffPS, convincingly show that a perturbative solution exists, in the form of a multipole expansion, with nice convergence properties in the space away from the brane profiles. Although it is possible to give heuristic arguments that the solution analytically continues to be a global solution of the nonlinear equation \Mongesource, a proof of this conjecture is still an open question.}.

The next step is to perform type IIB S-duality. The $D5$-branes turn into $NS5$-branes, and the new (unprimed) geometry is given in terms of the old one by
\eqn\Sduality{ ds^2 = e^{-(\varphi'-\varphi_0')} (ds')^2~, \qquad  e^{\varphi} = e^{-\varphi'}~, \qquad B_{2} = C_2'~,}
and, as one would expect, all R-R fields vanish.  While \Sduality\ trivially implies $H_3 = F_3'$, what we really need is a relation between the magnetic duals, $H_7 = \star H_3$ and $F_7' = \star' F_3'$.  One needs to be careful to take into account that Hodge duality involves the metric, which transforms under S-duality (since we are in string frame); $\star' (\star)$ denotes the hodge dual with respect to the primed (unprimed) metric.  One finds
\eqn\SdualHseven{ \eqalign{ H_{7} & = e^{-2(\varphi'-\varphi_0')} F_7'~.}}
The S-dual geometry is then,
\eqn\NSfive{ \eqalign{ & ds^2  = - dt^2 + d{\bf x}_{3}^2 + 2 \KK_{a \bbar} dz d\zbar^\bbar  + e^{-3A} d{\bf y}_{2}^2~, \cr
& e^{\varphi-\varphi_0} = e^{-3A/2}~, \cr
& H_7 = i e^{-3A} dt \wedge d( e^{3A} \KK_{a \bbar} dz^a \wedge d\zbar^{\bbar} ) \wedge d^3 {\bf x}~,}}
where $H_7 = \star dB_2$. Equations \gAK, \fluxeom\ determine $\KK_{a\bbar}$, $e^{-3A}$ as before.  The asymptotic string coupling is $g_s = e^{\varphi_0} (= 1/g_{s}')$.

Later we will require an expression for $B_2$ directly.  After some algebraic gymnastics (see Appendix A for details), one finds that $H_3 = \star H_7$ takes the form
\eqn\NSHthree{ H_3 = - i \left( \d_a e^{-3A} dz^a - \delbar_{\abar} e^{-3A} d\zbar^\abar \right) \wedge d^2 {\bf y} - i \epsilon_{i}^{\phantom{i}j} \d_j \KK_{a\bbar} dy^i dz^a d\zbar^\bbar~,}
where $i,j$ run over the directions ${\bf y} = y^i = (x^6,x^7)$.  Using \fluxsource, one finds that $dH_3$ is delta-function localised on the brane web. As a consequence, $H_3$ may only be trivialised away from the brane locus; in Appendix A we discuss in detail how to construct such a trivialisation $dB_2 = H_3$, with the result being
\eqn\Bfinal{ \eqalign{&  B_{2} = \frac{i}{2} \epsilon_{i}^{\phantom{i}j} \left( \KK_{ja}^{\rm reg.} dz^a - \KK_{j\bar a}^{\rm reg.} d\zbar^\abar \right) dy^i~, \quad {\rm with} \cr
& \KK_{ja}^{\rm reg.} \equiv  \KK_{ja} - \d_j {\tilde f}({\bf y}) \d_a \log{(\lambda - \lambda_0)}~,}}
where ${\tilde f}$ is a solution to
\eqn\singulargt{ \Delta_{\bf y} {\tilde f} = - \frac{Q_0}{2\pi} \delta^{(2)}({\bf y} - {\bf y}_0)~.}
For notational convenience we have defined a function $\KK^{\rm reg.}$ related to $\KK$ by \Bfinal\ whose presence is necessary in order to correctly describe $H_3$ in the plane ${\bf y}={\bf y_0}$.

The geometry \NSfive\ describes intersecting $NS5$-branes in type IIB, but one can trivially T-dualise along, say,  $x^3$ to get intersecting $NS5$-branes in type IIA, with the expressions \NSfive--\Bfinal\ unchanged.

To summarise, \NSfive--\Bfinal, combined with \Mongesource, describe the supergravity background for a web of $NS5$-branes intersecting along flat four-dimensional Minkowski space $\IR^{1,3}$. The branes follow a holomorphic profile $\lambda(z) = \lambda_0$ in $(z^1,z^2)$ with the directions ${\bf y} = (x^6,x^7)$  orthogonal to the web.  This result is consistent with those in \FayyazuddinZU.

An interesting property of \Mongesource\ is that it is invariant under transformations of the form
\eqn\KKahlertrans{ \KK \rightarrow \KK' = \KK + H(y^i) \left[ f(z^a) + \fbar(\zbar^\abar) \right]~,}
where $H$ is a harmonic function on $\IR_{\bf y}^2$ satisfying $\Delta_{\bf y} H = 0$.  This transformation has no effect on the metric, but it does transform the $B$-field, giving rise to
\eqn\Bgt{ B_2 \rightarrow B_{2}' = B_{2} + \delta B_{2}, \quad {\rm with}\quad \delta B_{2} = ( \star_{\bf y} d_{\bf y} H ) \wedge {\rm Im}( \d_a f dz^a )~.}
Here $\star_{\bf y}$ and $d_{\bf y}$ denote the Hodge dual and exterior derivative on $\IR_{\bf y}^2$.  However, $\delta B_2$ is an exact form, not affecting the physical field strength. To see this note that ${\rm Im}(\d_z f dz^a) = d {\rm Im} f$, where $d$ is the exterior derivative on the full six-dimensional (or ten-dimensional) space, and that $d_{\bf y} \star_{\bf y} d_{\bf y} H = 0$ giving
\eqn\Bgttwo{ \delta B_2 = -d \left[ {\rm Im} f ( \star_{\bf y} d_{\bf y} H ) \right] \equiv d \Lambda~.}
%

\subsec{T-dualising $NS5$-branes into metric}

We wish to T-dualise \NSfive--\Bfinal\ along the $x^7$ direction, using the Buscher rules \refs{\BuscherSK,\BergshoeffAS}.  To this end, we take $x^7$ to be a circular direction with asymptotic periodicity $x^7 \sim x^7 +2\pi R_7$, and smear the supergravity fields describing the $NS5$-branes along the $x^7$ direction to create an isometry.  At this point it is convenient to denote the remaining overall transverse direction as $x^6 = y$.  Denoting the T-dualised fields with tildes, the Buscher rules applied to the configuration \NSfive--\Bfinal\ imply
\eqn\Tduality{ \eqalign{ {\tilde G}_{MN} = & G_{MN} + \frac{B_{7M} B_{7N} }{G_{77}}~, \qquad {\tilde G}_{7M} = \frac{B_{7M}}{G_{77}}~, \qquad {\tilde G}_{77} = \frac{1}{G_{77}}~, \cr
{\tilde B}_{MN} = & B_{MN}=0~, \qquad {\tilde B}_{M 7} = 0~, \qquad e^{{\tilde \varphi}} = \frac{e^{\varphi}}{\sqrt{G_{77}}} = \frac{\ell_s}{R_7} g_s \equiv {\tilde g}_s~, } \qquad (M,N \neq 7)~.}
Since $\KK$ is independent of $x^7$, it follows from \Bfinal\ that $B_2$ must always have a leg along $x^7$ implying $\tilde{B}_2 = 0$.  Furthermore, we see that the dilaton is constant.  As expected, the $NS5$-branes have dualised into pure geometry.  For an asymptotically flat solution such that $e^{-3A} \to 1$, we have $G_{77} \to R_7/\ell_s$.  \Tduality\ implies ${\tilde G}_{77} = \ell_s/R_7$, and identifying this with ${\tilde R}_7/\ell_s$, we get the usual relation ${\tilde R}_7 = \ell_{s}^2/R_7$.

Using \NSfive, \Bfinal, the dual metric is given by (after some gentle massaging)
\eqn\gcnice{ \eqalign{  & d{\tilde s}^2 = -dt^2 + d{\bf x}_{3}^2 + d{\tilde s}_{6}^2~, \qquad {\rm with} \cr
& d{\tilde s}_{6}^2 = e^{-3 A}  dy^2 + 2 \KK_{a\bbar} dz^a d\zbar^\bbar  + e^{3 A} \left[ d{\tilde x}^7 - \frac{i}{2} \left( \KK_{ya}^{\rm reg.} dz^a - \KK_{y \bbar}^{\rm reg.} d\zbar^\bbar \right) \right]^2~,}}
where $d{\tilde x}_7 = {\tilde R}_7 d\phi_7$.  The equations determining the warp factor and potential are
\eqn\Mongenice{ \eqalign{ &  e^{-3 A} = 4 \det{( \d \delbar \KK)}~, \qquad \Delta_{\bf y} \KK + 2 e^{-3A} = - \frac{{\tilde Q}}{2\pi} \delta(y-y_0) \log{|\lambda - \lambda_0|^2}~, \cr
\Rightarrow \quad &  \Delta_{\bf y} \KK + 8 \det{( \del \delbar \KK)}   = - \frac{{\tilde Q}}{2\pi} \delta(y-y_0) \log{|\lambda - \lambda_0|^2}~.}}
Here we have integrated \fluxsource, \Mongesource\ with respect to $z^a, \zbar^\bbar$ and chosen $\KK$ such that there are no holomorphic or anti-holomorphic pieces.  We have also introduced ${\tilde Q}_0 = Q_0/(2\pi R_7)$, the charge density of the smeared brane.  For a single $NS5$-brane, $Q_0 = 2(2\pi^2 \ell_{s}^2)$ and therefore ${\tilde Q}_0 = 2\pi {\tilde R}_7$.   Note that, away from the sources one has $e^{-3A} = - \frac{1}{2} \KK_{yy}$, where $\KK_{yy} \equiv \d_{y}^2 \KK$, but at the locations of the sources these quantities differ and it is $e^{-3A}$ that appears in the metric \gcnice.

The metric \gcnice--\Mongenice\ describes the T-dual geometry of the $NS5$-brane web characterized by the source terms in \Mongenice.  Although $\KK$ is singular at the locations of sources, we will see that the space \gcnice\ can be perfectly smooth.

Let us consider the role of the transformations, \KKahlertrans.  First, in order to preserve the isometry in $x^7$ used for the T-duality, we should restrict to a subset of transformations that are independent of $x^7$.  In this case there are only two linearly independent harmonic functions to consider, which we may take as $H(y) = 1,y$.  The most general transformation in this restricted class takes the form
\eqn\KKahlertransres{ \KK \rightarrow \KK' = \KK + y \left[ f_1(z^a) + \fbar_1(\zbar^\abar) \right] + f_{0}(z^a) + \fbar_0(\zbar^\abar) ~.}
There is no $B$-field in the dual geometry, \gcnice, but now the $f_1$ terms in $\delta \KK$ will show up in the metric.  Since $\frac{i}{2} \left( \delta \KK_{y a} dz^a - \delta \KK_{y\abar} d\zbar^\abar \right) = - d ( {\rm Im} f_1)$, the transformation induces a change of coordinates
\eqn\inducedshift{ {\tilde x}^7 \rightarrow {\tilde x}^7{}' = {\tilde x}^7 + {\rm Im} f_1~.}
Said differently, the metric will be invariant under \KKahlertransres\ provided we simultaneously shift ${\tilde x}^7$ to cancel \inducedshift.  Gluing together coordinate patches with such transitions on the overlaps can lead to nontrivial circle fibrations in the T-dual geometry, a familiar fact from the example of Taub-NUT.

Finally, let us remark on the supersymmetry of \gcnice, \Mongenice. At the level of supergravity, T-duality does not always preserve supersymmetry. A canonical example is to T-dualise the polar direction in a plane embedded in flat space. The initial background, being flat space, preserves all supersymmetries. Yet after applying the Buscher rules, one finds a background that is not supersymmetric. A sufficient condition for the preservation of supersymmetry is that the Killing spinors of the original solution be independent of the coordinate parameterising the T-duality direction \refs{\BakasBA,\AlvarezZR,\AlvarezAI}.  Since we have smeared the original geometry along the T-duality direction, one certainly expects the Killing spinors to fulfill this criteria.  Nonetheless, it is rewarding to explicitly check that \gcnice, \Mongenice\ is indeed $1/4$-BPS.  A summary of this calculation is given in Appendix B.  For a convenient choice of local frame we find that, remarkably, the Killing spinors are independent of all coordinates.

Our strategy now will be to find solutions to \Mongenice\ by demanding the equivalence of \gcnice\ to known metrics for the duals of particular configurations of $NS5$-branes. As a warm up, we show in Appendix C how to map \gcnice\ to Taub-NUT, thus finding the solution for $\KK$ corresponding to an arbitrary number of parallel $NS5$-branes. A more interesting example however is the conifold, which we turn to in the next section. We will find a solution $\KK$ and a coordinate transformation that maps \gcnice\ to the conifold metric of \CandelasJS.  The transformation is nontrivial and leads to some interesting observations regarding the duality relation between the $NS5$-branes and the conifold.

\newsec{Using the conifold to determine $NS5$-brane metrics}

In this section we determine the explicit supergravity solution for a special case of the brane webs discussed above, where we have a single $NS5$-brane stretched in $z^1$ and located at $y = z^2 = 0$, and a second $NS5$-brane stretched in $z^2$ and located at $y = z^1 = 0$.  Following standard nomenclature, we will refer to these branes as the $NS$-brane and the $NS'$-brane respectively.  These branes also span $\IR^{1,3}$ labeled by $(t,{\bf x})$ and are smeared on the $x^7$ circle. The background for this configuration is written down in \NSfive--\Bfinal\ up to an undetermined function $\KK$. Following the general procedure outlined in the previous section, we T-dualise along the $x^7$ circle resulting in \gcnice, but this time we can make use of the well known result of \BershadskySP\ that this geometry is to be identified with the conifold Calabi-Yau geometry. As this metric is known (see \CandelasJS\ for more details), we can use this to determine the function $\KK$, show that it is a solution to \Mongenice, and that it has the relevant properties appropriate for a pair of intersecting $NS5$-branes. In this way, we will obtain the explicit 1/4-BPS background \NSfive--\Bfinal\ for a pair of intersecting $NS5$-branes.

In the first subsection we will review some pertinent facts about the conifold, largely following \CandelasJS. Next we will determine a local solution to the source-free Monge-Ampere equation with the aid of a probe brane. We will then show how to extend this solution globally, showing that it solves the sourced Monge-Ampere equation \Mongenice\ with the appropriate singularity structure. We tie these results together by writing down the complete solution for a pair of intersecting $NS5$-branes, smeared in a single transverse direction.

\subsec{A lightening review of the conifold}

The conifold is a non-compact Calabi-Yau hypersurface. It may be described algebraically as the vanishing of a polynomial in $\IC^4$
\eqn\xuvw{ x u = v w~,}
with $x,u,v,w$ the coordinates of $\IC^4$.  When $x\ne0$ we may solve \xuvw\ for $u$ and use $\zeta^\alpha = (x,v,w)$ as the independent complex coordinates. The large isometry group of the conifold implies that the K\"ahler potential, $\FF$, can only be a function of the radial coordinate
\eqn\rdef{ r^2 =  |x|^2 +  |v|^2 + |w|^2 + |u|^2 = \frac{( |x|^2 + |v|^2 )(|x|^2 + |w|^2)}{|x|^2}~.}
In the last step we restricted to the patch $x\ne 0$, such that $u = v w/x$.  The K\"ahler metric is
\eqn\kahmetone{ \eqalign{ & ds_{con}^2 = 2 g_{\alpha {\bar\beta}} d\zeta^\alpha d{\bar\zeta}^{\bar\beta}~, \qquad {\rm with} \cr
&  g_{\alpha{\bar\beta}} = \del_\alpha \delbar_{\bar\beta} \FF = (\del_\alpha \delbar_{\bar\beta} r^2) \FF' + (\del_\alpha r^2) (\delbar_{\bar\beta} r^2) \FF''~,}}
where the prime denotes differentiation with respect to $r^2$. Using \rdef, one finds that the determinant $g \equiv \det{(g_{\alpha{\bar\beta}})}$ takes a simple form.   Ricci flatness, $R_{\alpha{\bar\beta}} = \del_\alpha \delbar_{\bar\beta} \log{g}$ = 0, reduces to the following condition:
\eqn\Rflatcon{ (\gamma^3 )' = 2 L^2 r^2 \qquad {\rm with} \quad \gamma \equiv r^2 \FF'~.}
One can solve this condition to get an explicit expression for the K\"ahler potential:
\eqn\Fresult{ \FF = \frac{3}{2} L^{2/3} r^{4/3}~. }
Plugging this result into \kahmetone, with $r^2$ given by \rdef, yields a Ricci flat metric on the conifold.  The length scale $L$ is an integration constant.  Since we take our coordinates $(x,v,w)$ to have dimensions of length, $L$ guarantees that the components of the metric, $g_{\alpha{\bar\beta}}$,  are dimensionless.  In the following, when we show how to rewrite the conifold metric in terms of brane variables in the form \gcnice, we will see that $L$ may be related to the asymptotic radius of the ${\tilde x}^7$ circle, ${\tilde R}_7$.  One may recover the results of \CandelasJS\ by setting $L=1$.

In the neighborhood of $x = 0$, one should use a different coordinate system.  If $u$ is nonzero we may solve \xuvw\ for $x$ and take $\zeta^\alpha = (u,v,w)$ as complex coordinates.  The K\"ahler potential is again \Fresult, where $r^2$ is given in terms of $(u,v,w)$ by making the replacement $x \to u$ at the end of \rdef.

By introducing a pair of Euler angles $\theta_i,\phi_i,\psi_i$, $i=1,2$, the metric may be written as a cone over the Einstein space $T^{1,1} = \frac{SU(2) \times SU(2)}{U(1)}$ via the coordinate transformation
\eqn\changeofvar{ \eqalign{ & x = r \cos{\frac{\theta_1}{2}} \cos{\frac{\theta_2}{2}} e^{\half i (\psi + \phi_1 + \phi_2)}~, \cr
& v = - r \cos{\frac{\theta_1}{2}} \sin{\frac{\theta_2}{2}} e^{\half i(\psi + \phi_1 - \phi_2)}~, \cr
&  w = r \sin{\frac{\theta_1}{2}} \cos{\frac{\theta_2}{2}} e^{\half i(\psi - \phi_1 + \phi_2)}~, \cr
& u = -r \sin{\frac{\theta_1}{2}} \sin{\frac{\theta_2}{2}} e^{\half i(\psi - \phi_1 - \phi_2)}~,}}
where $\psi \equiv \psi_1 + \psi_2$ generates a $U(1)$ embedded symmetrically in $SU(2)\times SU(2)$.  After changing the radial variable to $\rho$, given by
\eqn\rhodef{ \rho \equiv \sqrt{3} L^{1/3} r^{2/3}~,}
one finds that \kahmetone\ takes the form
\eqn\conemet{ds_{con}^2 = d\rho^2 + \rho^2 \left( \frac{1}{6} \sum_{i=1}^{2} (d\theta_{i}^2 + \sin^2{\theta_i} d\phi_{i}^2 ) + \frac{1}{9} \left( d\psi +\sum_{i=1}^2 \cos{\theta_i} d\phi_i \right)^2 \right)~.}
The space $T^{1,1}$ is topologically $S^2\times S^3$ and \conemet\ is Ricci-flat everywhere except at the singularity $\rho=0$. We will now make use of these results to determine $\KK$ in \gcnice.

\subsec{Local analysis via a probe brane}

Our goal in this section is to determine a change of variables and a function $\KK$ such that the metric $d{\tilde s}_{6}^2$, \gcnice, is equivalent to the conifold metric, \kahmetone. Firstly, we identify the conifold coordinates $v,w$ in \xuvw\ with the brane variables $z^1,z^2$ in \gcnice:
\eqn\vwzrelation{ (v,w) = (z^1, z^2)~.}
Secondly, we need to determine a change of variables
\eqn\yxsevencov{ y = y(x,\xbar,v,\vbar,w,\wbar)~, \qquad {\tilde x}^7 = {\tilde x}^7(x,\xbar,v,\vbar,w,\wbar)~.}
This requires a little bit of work and a useful tool turns out to be the DBI action of a probe $D5$-brane extended in $\IR^{1,3}$ and the $x,\xbar$ plane, or equivalently, the $y$, ${\tilde x}^7$ cylinder. The $D5$-brane is sitting at an arbitrary point $(v,w)$ in $\IC_{v} \times \IC_{w}$ and we consider spacetime fluctuations in the $w$ direction, letting $w = w(x^\mu)$.  Using the pullback of the conifold metric \kahmetone, a short calculation of the DBI action gives\foot{We could have just as easily turned on fluctuations in the $v$-direction as there is a symmetry interchanging $v$ and $w$. However, studying just the $w$ fluctuations turns out to be sufficient for our purposes.}
\eqn\DBIconifold{ \LL_{D5} = -T_5 \int dx d\xbar \sqrt{ - g_{x\xbar}^2 } \left\{ 1 + \left( g_{w\wbar} - \frac{ |g_{w\xbar}|^2 }{g_{x\xbar}} \right) |\d_\mu w|^2 + \cdots \right\}~,}
On the other hand, if we use the pullback of the metric that naturally appears from dualising the brane system, \gcnice, one finds, after some work and some nontrivial cancellations,
\eqn\DBIbrane{ \LL_{D5} = -T_5 \int dy d{\tilde x}^7 \left\{ 1 + \KK_{w\wbar} |\d_\mu w|^2 + \cdots \right\}~.}
The ellipses represent higher derivative terms in the expansion of the DBI action; the terms displayed will be sufficient for our purposes.

These two Lagrangians must give equivalent descriptions of the probe brane and the scalar field $w(x^\mu)$.  By equating the energy densities when the scalar field is set to zero we learn that the Jacobian restricted to the $x$-$\xbar$ sector for the change of variables \yxsevencov\ must satisfy
\eqn\jacfix{ \del_x y  \delbar_\xbar {\tilde x}^7 - \delbar_\xbar y \del_x {\tilde x}^7 = \pm i g_{x\xbar}~.}
To proceed we make an ansatz for how $y$, ${\tilde x}^7$ depend on $x$.  Given that ${\tilde x}^7$ is an isometry direction, a natural guess is that ${\tilde x}^7$ depends only on the phase of $x$, while $y$ depends only on the magnitude:
\eqn\xdependence{ y = y(|x|)~, \qquad {\tilde x}^7 = {\tilde x}^7(\phi_x)~, \qquad {\rm where} \quad \phi_x = - \frac{i}{2} \log{\left(x/\xbar \right)}~.}
Note that $y, {\tilde x}^7$ may still depend on $v,w$---we are omitting their functional dependence here for notational simplicity.  Given \xdependence, \jacfix\ simplifies to
\eqn\jacsimp{ (\del_{|x|^2} y)( \del_{\phi_x} {\tilde x}^7) = \pm g_{x\xbar}~.}
Since $\FF = \FF(|x|^2)$, we have that $g_{x\xbar} = \del_{|x|^2} (|x|^2 \del_{|x|^2} \FF)$ is a function of $|x|$ and not $\phi_x$.  Thus \jacsimp\ implies $\del_{\phi_x} {\tilde x}^7$ must be independent of $\phi_x$.  Let $\del_{\phi_x} {\tilde x}^7 = c$, where $c$ may depend on $v,w$.  Plugging this into \jacsimp, we integrate both sides with respect to $|x|^2$, obtaining
\eqn\yresult{ \eqalign{ \del_{\phi_x} {\tilde x}^7 =&~ c \cr
 \Rightarrow \quad c y + d =&~ \pm |x|^2 \d_{|x|^2} \FF = \pm \frac{ L^{2/3} (|x|^4 - |v|^2 |w|^2) }{|x|^{4/3} (|x|^2 + |v|^2)^{1/3} (|x|^2 + |w|^2)^{1/3} }~,}}
where $d$ is another undetermined function of $v,w$.  The choice of sign, as well as the domain of $y$ in which \yresult\ applies, are global issues that we will have to deal with shortly, but for now we leave this data unspecified and continue with our local analysis.

We have already gathered sufficient information to determine the warp factor, $e^{-3A}$.  Given the identification \vwzrelation, the change of coordinates formula relating the metrics \gcnice\ and \kahmetone\ takes a relatively simple form in the $x$-$\xbar$ block:
\eqn\gJac{ \left( \matrix{ \del_x y & \del_x {\tilde x}^7 \cr \delbar_\xbar y & \delbar_\xbar {\tilde x}^7 } \right) \left( \matrix{ e^{-3A} & 0 \cr 0 & e^{3A} } \right) \left( \matrix{ \del_x y & \delbar_\xbar y \cr \del_x {\tilde x}^7 &  \delbar_\xbar {\tilde x}^7 } \right) = \left( \matrix{ 0 & g_{x \xbar} \cr g_{x \xbar} & 0 } \right)~.}
Equation \jacfix\ is reproduced if we take the square root of the determinant of each side, but now we want to consider the individual component equations.  Using \yresult, we find that \gJac\ is satisfied if and only if
\eqn\Kyy{ e^{-3A} = \frac{c^2}{2 |x|^2 g_{x\xbar}}  =  \frac{3 c^2  |x|^{4/3} ( |x|^2 + |v|^2 )^{4/3} (|x|^2 + |w|^2 )^{4/3} }{2 L^{2/3} \Delta}~,}
where $\Delta$ is defined by
\eqn\Deltadef{ \Delta \equiv 2 |x|^8 + 3 |x|^6 (|v|^2 + |w|^2) + 8 |x|^4 |v|^2 |w|^2 + 3 |x|^2 |v|^2 |w|^2 (|v|^2 + |w|^2) + 2 |v|^4 |w|^4~.}
Note that $e^{-3A}, \KK$ are defined as a functions of $(y,z^a,\zbar^\abar)$, or equivalently $(y,v,w,\vbar,\wbar)$.  In expressions such as \Kyy, $|x|^2$ must be viewed as a function of $(y,v,w,\vbar,\wbar)$, obtained by solving \yresult.

Now consider the quadratic terms in \DBIconifold\ and \DBIbrane.  Given the equivalence of the volume elements, it must be that
\eqn\Kwwbar{ \KK_{w \wbar} = g_{w \wbar} - \frac{|g_{w \xbar}|^2}{g_{x\xbar}}~.}
Turning on fluctuations in the $v$ direction instead leads to
\eqn\Kvvbar{ \KK_{v \vbar} = g_{v \vbar} - \frac{|g_{v \xbar}|^2}{g_{x\xbar}}~.}
Cranking the handle and computing the metric $g_{\alpha{\bar\beta}}$ means we can determine
\eqn\Kwwbarvvbar{ \eqalign{ & \KK_{v\vbar} = \frac{ L^{2/3} |x|^{2/3} (|x|^2 + |v|^2)^{2/3} (|x|^2 + |w|^2)^{2/3} }{\Delta} (2 |x|^4 + 3 |x|^2 |w|^2 + 2 |v|^2 |w|^2)~, \cr
&  \KK_{w\wbar} = \frac{ L^{2/3} |x|^{2/3} (|x|^2 + |v|^2)^{2/3} (|x|^2 + |w|^2)^{2/3} }{\Delta} (2 |x|^4 + 3 |x|^2 |v|^2 + 2 |v|^2 |w|^2)~.}}

Suppose that we are not in the plane of the sources, \ie\ suppose $y \neq y_0$.  Then \Kyy\ determines $\KK_{yy}$ through $e^{-3A} = -\frac{1}{2} \KK_{yy}$, and with $\KK_{yy},\KK_{v\vbar},\KK_{w\wbar}$ in hand, we use equation \Monge\ to solve for $\KK_{v \wbar}$, finding
\eqn\Kvwbar{ \KK_{v\wbar} = \frac{L^{2/3} |x|^{8/3} (|x|^2 + |v|^2)^{2/3} (|x|^2 + |w|^2)^{2/3} }{\Delta} \vbar w \sqrt{ 1 + \left( 8 - \frac{3 c^2}{2 L^2} \right) \frac{\Delta}{|x|^4 |v|^2 |w|^2} }~.}
This guarantees that we have solved the Monge-Ampere equation, \Monge, as an algebraic equation for unrelated functions $\KK_{mn}$, where the indices run over $m,n = y,a,\abar$.  As a differential equation, of course, we must require that $\KK_{mn}$ be partial derivatives of a function $\KK$. A necessary consequence of this is that the functions $\KK_{mn}$ must satisfy an integrability condition $\d_p \KK_{mn} = \d_m \KK_{np} = \d_n \KK_{pm}$. This holds if and only if
\eqn\cdfix{ c = \pm \frac{4 L}{\sqrt{3}}~, \qquad {\rm and} \quad  d = const \equiv -c y_0~.}
In showing \cdfix, $|x|^2$ is to be viewed as a function of $(y,v,w,\vbar,\wbar)$ as implicitly determined by \yresult.  While \yresult\ is not invertible, derivatives of \yresult\ may be taken implicitly and expressions for $\d_m |x|^2$ are easily obtained.  The sign choice in \cdfix\ simply determines the orientation of the ${\tilde x}^7$ coordinate relative to $\phi_x$; we will choose the plus sign in the following.   Note that this choice is independent of the one in \jacsimp.  The identification of $d$ with $y_0$, the $y$-location of the sources, will be justified in the next subsection. In the meantime we choose our coordinate system such that $y_0 = 0$.

The integrability condition $\d_p \KK_{mn} = \d_m \KK_{np} = \d_n \KK_{pm}$, though a necessary condition for the existence of a solution to \Monge, is not sufficient.  To demonstrate that \Monge\ actually has a solution, we will construct the function $\KK$ explicitly. Starting from the expression \Kyy, we see from \jacsimp\ and \yresult\ that $g_{x\xbar} = \pm c \d_{|x|^2} y$, which together with identity $\d_{|x|^2} y = ( \d_y |x|^2 )^{-1}$, implies that (away from the sources):
\eqn\Kyytwo{ \KK_{yy} =   \mp \frac{c}{|x|^2} \del_y |x|^2~.}
The sign here is correlated with the one in \yresult.  Let ${\hat \KK}$ be a function such that $\del_{y}^2 {\hat \KK} = \KK_{yy}$.  Integrating both sides of \Kyytwo\ with respect to $y$, holding $v,w$ fixed, gives
\eqn\Khaty{ {\hat \KK}_y = \mp c \log{\frac{|x|^2}{c^2}} + f_{1}(v,w,\vbar,\wbar)~,}
where ${\hat \KK}_y \equiv \del_y {\hat \KK}$ and $f_1$ is an arbitrary function of $v,w$.  We can integrate again, using the relation $dy = (\del_{|x|^2} y) d|x|^2 = \pm \frac{1}{c} g_{x\xbar} d|x|^2$, for fixed $v,w$.  Recalling that $g_{x\xbar} = \del_{|x|^2} (|x|^2 \del_{|x|^2} \FF)$, we get
\eqn\Khat{ \eqalign{ {\hat \KK} =&~ - \int d|x|^2 \left[ \log{\frac{|x|^2}{c^2}} \del_{|x|^2} (|x|^2 \del_{|x|^2} \FF) \right] + y f_1(v,w,\vbar,\wbar) + f_0(v,w,\vbar,\wbar) \cr
=&~ \FF - \log{\frac{|x|^2}{c^2}} \left( |x|^2 \del_{|x|^2} \FF \right) +  y f_1(v,w,\vbar,\wbar) + f_0(v,w,\vbar,\wbar) \cr
=&~ \FF \mp c y \log{\frac{|x|^2}{c^2}} +  y f_1(v,w,\vbar,\wbar) + f_0(v,w,\vbar,\wbar)~.}}

We may now ask whether or not $\del_a \delbar_\bbar {\hat \KK} = \KK_{a\bbar}$, where the $\KK_{a\bbar}$ are given in \Kwwbarvvbar\ and \Kvwbar\ with $c^2 = 16 L^2/3$.  Remarkably, the equality holds, provided the $f_0,f_1$ terms give a vanishing contribution---that is, provided they are sums of holomorphic and anti-holomorphic functions.  Imposing this restriction, along with the constraint that $\KK$ be real, we conclude that
\eqn\Kgeneral{ \KK(y,|v|,|w|) = \FF \mp c y \log{\frac{|x|^2}{c^2}} + y(f_1(v,w) + \fbar_1(\vbar,\wbar)) + f_{0}(v,w) + \fbar_0(\vbar,\wbar),~}
is a solution to the source free Monge-Ampere equation \Monge\ (keeping in mind that $|x|$ is implicitly determined in terms of $y,|v|,|w|$ via \yresult).

Partials of the form $\KK_{ya}, \KK_{y\abar}$ computed from \Kgeneral\ are clearly nonzero.  We find
\eqn\Kyvwexplicit{ \eqalign{ \KK_{yv} &= \mp \frac{c \vbar (|x|^2 + |w|^2 ) ( |x|^4 + 3 |x|^2 |w|^2 + 2 |v|^2 |w|^2 )}{\Delta} + \d_v f_1~,\cr
 \KK_{yw} &= \mp \frac{c \wbar (|x|^2 + |v|^2 ) ( |x|^4 + 3 |x|^2 |v|^2 + 2 |v|^2 |w|^2 )}{\Delta} + \d_w f_1~,}}
with the $\KK_{y\abar}$ given by conjugation.  These partials do not enter in the Monge-Ampere equation, but they do enter into the determination of the metric \gcnice.  In particular, the appearance of the arbitrary holomorphic function $f_1$ may seem disturbing, as we expect the metric  to be unique.  However, we have yet to fully specify the change of variables that maps \gcnice\ into the conifold, \kahmetone.  From \yresult\ we have that
\eqn\yxsevengen{c y = \pm |x|^2 \d_{|x|^2} \FF~, \qquad  {\tilde x}^7 = c \phi_x + h(v,w,\vbar,\wbar)~,}
where $c = \frac{4 L}{\sqrt{3}}$ and $h$ is an undetermined function.  Explicit computation shows that the metric \gcnice, with $\KK$ given by \Kgeneral\ and change of variables given by \yxsevengen, matches the conifold metric \kahmetone, if and only if
\eqn\hfix{ h(v,w,\vbar,\wbar) = \frac{i}{2} \left( f_1(v,w) - \fbar_1(\vbar,\wbar) \right) = -{\rm Im}(f_1(v,w))~.}
In this case the $f_1$ contribution to $d{\tilde x}^7$ precisely cancels the $f_1$ contribution to $\frac{i}{2} ( \KK_{y a} dz^a - \KK_{y\abar} d\zbar^\abar)$ in \gcnice.  (Since we are working away from $y = y_0$, there is no difference between $\KK_{ya}^{\rm reg.}$ and $\KK_{ya}$).  This is consistent with our discussion around equation \inducedshift.  We recognize the $f_1,f_0$ terms in \Kgeneral\ as K\"ahler transformations of the sort \KKahlertransres.  They induce a shift in the coordinate ${\tilde x}^7$, which is offset by \hfix.

In summary, the coordinate transformation
\eqn\coordtrans{ \eqalign{ & y = \pm \frac{1}{c} |x|^2 \d_{|x|^2} \FF = \pm \frac{ \sqrt{3} (|x|^4 - |v|^2 |w|^2) }{4 L^{1/3} |x|^{4/3} (|x|^2 + |v|^2)^{1/3} (|x|^2 + |w|^2)^{1/3}}~, \cr
& (z^1,z^2) = (v,w)~, \qquad {\tilde x}^7 = \frac{4 L}{\sqrt{3}} {\rm Im} \left[ \log{x} - f_1(v,w) \right]~,}}
maps \gcnice\ into the conifold, \kahmetone.  The potential $\KK$ is given by
\eqn\Ksingular{ \eqalign{ \KK(y,|v|,|w|) =&~ \KK^0(y,|v|,|w|) + 2 y {\rm Re}(f_1(v,w)) + 2 {\rm Re}(f_0(v,w))~, \cr
 \KK^0(y,|v|,|w|) \equiv&~ \FF - \frac{4 L}{\sqrt{3}} \left(|x|^2 \d_{|x|^2} \FF \right) \log{\frac{|x|^2}{c^2}}  \cr
 =&~L^{2/3} \left[  \frac{ 3(|x|^2 + |v|^2)(|x|^2 + |w|^2) - 2 (|x|^4 - |v|^2 |w|^2) \log{\frac{|x|^2}{c^2}} }{ 2 |x|^{4/3} (|x|^2 + |v|^2)^{1/3} (|x|^2 + |w|^2)^{1/3} } \right]~.}}

Some comments are in order.
\item{(1)} We have been working on a coordinate patch of the conifold defined by $x \neq 0$, as we solved $x u = v w$ for $u$ by dividing through by $x$.  A related fact is that we have been solving the source free Monge-Ampere equation \Monge.  Recall that the sources are located at $y = v = 0$ or $y = w = 0$.  In \cdfix\ we identified the integration constant $d$ with $y_0=0$, implying that the brane locus is contained in the hypersurface $|x|^2 = |v| |w|$, and hence $x=0$ at the source locations. Since this is outside of our coordinate patch, a solution to the source free equation is also a solution to the sourced equation.  However, the identification $d \propto y_0$ remains to be justified; doing so will require constructing a global solution for $\KK$ and studying it's behavior as one approaches the sources.
\item{(2)}  The definition of the coordinate $y$ is ambiguous and incomplete.  For $|x|^2 \in (0,\infty)$ with $v,w \neq 0$, we have $y \in (-\infty,\infty)$, but when $v$ or $w=0$, $y \in (0,\infty)$.  However, from the brane picture we expect $y \in (-\infty,\infty)$ for any $v,w$, and $\IR_{y} \times \IC_{v} \times \IC_{w}$, along with the ${\tilde x}^7$ circle, should cover the entire conifold.  Clearly the missing $y$'s are related to the fact that when $v$ or $w = 0$, we can have $x = 0$ with $u$ arbitrary.  In other words, a complete definition of $y$ must make use of a second patch, containing $x =0$, where $(u,v,w)$ are the good coordinates.
\item{(3)}  The physical ambiguity in ${\tilde x}^7$ corresponds the the freedom to choose asymptotically nontrivial $f_1$.  We will see that this freedom is essential in the global construction, and is related to the inability to trivialise the ${\tilde x}^7$ circle over the space $(\IR_y \times \IC_v \times \IC_w)$ minus the brane locus.
\item{(4)}  There is additional ambiguity in $\KK$.  We are instructed to solve for $|x|^2$ as a function of $y$, but as can be seen from \coordtrans, this amounts to finding roots of a sextic polynomial.  It is easy to show that this polynomial always has at least one real positive root, and may have three.  In the next subsection we will determine the appropriate root in the limit one approaches the sources.  Later, when we write down the explicit geometry for the intersecting branes, we will circumvent the issue by using a different coordinate system.
\item{(5)} So far we have been focusing on the geometry side of the T-duality map.  Given that \Ksingular\ solves \Mongesource, we have also determined the supergravity background, \NSfive, for two intersecting $NS5$-branes smeared on a transverse circle.  We will study this background further in section $3.5$, but it is interesting to note that when plugged into \NSfive, the function \Ksingular\ does not yield an asymptotically flat geometry. This seems to imply the background we have constructed is not describing the full geometry of the intersecting $NS5$-brane system, but rather is its near horizon limit. Presumably there is a more general background that is asymptotically flat and describes the complete intersecting five-brane system.
\item{} A helpful analogy is the following. At the level of supergravity the T-dual of the near horizon limit of a stack of parallel $NS5$-branes is the Eguchi-Hanson space. This geometry is not asymptotically flat. On the other hand, the T-dual of the complete five-brane background smeared along the T-duality circle, as described by the smeared CHS solution \CallanAT, or ``H-monopole'' \BanksRJ, is the Taub-NUT metric. One can obtain Taub-NUT from Eguchi-Hanson by adding a constant to the harmonic function appearing in the Eguchi-Hanson metric. The Taub-NUT and Eguchi-Hanson metrics are related by this simple superposition due to the large amount of symmetry present; the Monge-Ampere equation becomes linear. In the example we are considering here, the $NS5$-branes are not parallel and the number of symmetries is reduced. Consequently, the Monge-Ampere equation that $\KK$ satisfies is a nonlinear PDE, and the superposition principle does not hold. It would be very interesting to determine the complete background describing intersecting five-branes with an asymptotically flat limit. Presumably though, the relation to the geometry we have determined here is rather nontrivial.
\item{(6)} It seems natural to identify the constant $c = \frac{4 L}{\sqrt{3}}$ with the asymptotic radius of the ${\tilde x}^7$ circle, ${\tilde R}_7$.  However, this identification is not obvious since the conifold is not asymptotically flat---the ${\tilde x}^7$ direction does not approach a circle of constant radius.  By studying the singularity structure of the solution near one brane and far from the other, and matching onto known results, we will be able to argue that indeed $c = {\tilde R}_7$.
\item{(7)} The relation between the potentials $\KK$, $\FF$, and the coordinates $y, |x|^2$ may be described as a Legendre transform.  Specifically, $c y$ and $\xi \equiv \log{(|x|^2/c^2)}$ are the dual variables.  From \coordtrans, \Ksingular\ we have
\eqn\legendre{ \KK(y) = \left\{ \FF(\xi) \mp c y \xi \right\}_{{\rm max} \ \xi} ~,  \qquad c y = \pm \d_{\xi} \FF~.}
In section $3.4$ we will use this relation to prove that, locally, any K\"ahler metric with a $U(1)$ isometry can be put in the form \gcnice, and that Ricci flatness of the K\"ahler manifold is equivalent to the Monge-Ampere equation \Monge.

\subsec{Global construction and singularity structure}

Suppose we had chosen to work with variables $\zeta^\alpha = (u,v,w)$ instead, valid on the patch $u \neq 0$.  Since $\FF$ is symmetric under the interchange $x \leftrightarrow u$, we could have derived formulae identical to \coordtrans, \Ksingular, but with $x \to u$.  Generically, one would expect these equations to define a different coordinate $y'$ in the second patch, with some coordinate transformation $y'(y)$ relating them on the overlap.  However, from the brane picture we expect that it should be possible to extend the coordinate $y$ so that it is well defined everywhere, and definitions in different patches should agree on the overlaps.  The following simple observation shows that this is possible.  For $v,w$ nonzero, implying $u,x$ nonzero, we may substitute $|x| = |v| |w|/|u|$ into \coordtrans, obtaining
\eqn\overlap{ \frac{  (|x|^4 - |v|^2 |w|^2) }{|x|^{4/3} (|x|^2 + |v|^2)^{1/3} (|x|^2 + |w|^2)^{1/3}} = - \frac{(|u|^4 - |v|^2 |w|^2) }{|u|^{4/3} (|u|^2 + |v|^2)^{1/3} (|u|^2 + |w|^2)^{1/3}}~.}
Therefore we define the coordinate $y$ on each patch via
\eqn\yglobal{ y = \left\{ \matrix{ \frac{ \sqrt{3} (|x|^4 - |v|^2 |w|^2) }{ 4 L^{1/3} |x|^{4/3} (|x|^2 + |v|^2)^{1/3} (|x|^2 + |w|^2)^{1/3}}~, & x \neq 0~, \cr
- \frac{\sqrt{3} (|u|^4 - |v|^2 |w|^2) }{ 4 L^{1/3} |u|^{4/3} (|u|^2 + |v|^2)^{1/3} (|u|^2 + |w|^2)^{1/3}}~, & u \neq 0~.} \right. }
This definition agrees on the overlap where both $x,u$ are nonzero, thanks to \overlap.  The region $v,w,u = 0$, $|x| >0$ is mapped to the positive $y$ axis while $v,w,x =0$, $|u|>0$ is mapped to the negative $y$-axis.  We will refer to these patches as the upper $(+)$ and lower $(-)$ patches respectively.  Note that if both $x = u = 0$, then since $v = 0$ or $w = 0$ (or both), and these points are contained in the hypersurface $y = 0$, they correspond precisely to the brane locus.  The upper and lower patches cover everything except the brane locus; we will discuss the structure of the solution as one approaches the sources momentarily.

Now consider the coordinate ${\tilde x}^7$.  Suppose on the upper patch we choose $f_{1}^{(+)} = 0$, so that ${\tilde x}^{7(+)} = \frac{4 L}{\sqrt{3}} \phi_x$.  This is well defined everywhere on the patch, since $x \neq 0$---that is, it gives a trivialisation of the ${\tilde x}^7$-circle on the upper patch.  Can we give a trivialisation on the lower patch that agrees on the overlap?  This would amount to a trivialisation of the circle on the full space $\IR_y \times \IC_v \times \IC_w$ minus the brane locus, but it is easy to see that this is not possible.  On the lower patch we will have the general solution ${\tilde x}^{7(-)} = \pm \frac{4 L}{\sqrt{3}} \phi_u - {\rm Im} f_{1}^{(-)}$.  (Note the sign choice in \cdfix\ can be made independently of the sign choice in \yresult).  On the overlap, the defining equation of the conifold implies $\phi_x + \phi_u = \phi_v + \phi_w$, so in order for our definitions of ${\tilde x}^7$ to agree, we should take the minus sign and choose $f_{1}^{(-)} = - \frac{4 L}{\sqrt{3}} (\log{v} + \log{w})$.  However, this definition is not well defined on the entire patch, which includes points where $v,w = 0$.  We conclude that it is not possible to trivialise the ${\tilde x}^7$-circle\foot{This should not be surprising since exactly the same phenomenon occurs in the Taub-NUT example.  (See Appendix C).}, and considering $f_{1}^{(+)} \neq 0$ does not alter the result.  Instead, we give a coordinatisation on each patch,
\eqn\xsevenglobal{ {\tilde x}^{7(+)} = \frac{4 L}{\sqrt{3}} \phi_x~, \qquad {\tilde x}^{7(-)} = - \frac{4 L}{\sqrt{3}} \phi_u~,}
such that on the overlap they are related by a coordinate transformation of the sort \inducedshift:
\eqn\xtransition{ {\tilde x}^{7(-)} = {\tilde x}^{7(+)} - \frac{4 L}{\sqrt{3}} {\rm Im}( \log{v} + \log{w} )~.}

Given \yglobal, \xsevenglobal, it follows from the analysis of the previous section that the function $\KK$ is given on each patch by
\eqn\Kglobal{ \eqalign{ & \KK^{(+)}(y) =  \left[ \FF(|x|^2) - \frac{4 L}{\sqrt{3}} y \log{\frac{|x|^2}{c^2}} \right]_{|x|^2 = |x|^2(y)} ~, \cr
& \KK^{(-)}(y) = \left[ \FF(|u|^2) + \frac{4 L}{\sqrt{3}} y \log{\frac{|u|^2}{c^2}} \right]_{|u|^2 = |u|^2(y)} ~,}}
where we have suppressed $(|v|,|w|)$ in the arguments of functions, $\FF$ is given by
\eqn\Fglobal{ \FF(|X|^2) = \frac{3 L^{2/3} (|X|^2 + |v|^2)^{2/3} (|X|^2 + |w|^2)^{2/3} }{2 |X|^{4/3} }~, \qquad X = x,u,}
and $|x|^2(y)$, $|u|^2(y)$ are determined by solving \yglobal.  Consider the difference between these expressions on the overlap.  Suppose we have root $|x| = |x_r|$ of \yglobal.  This corresponds to a root $u = |u_r|$, related on the overlap by $|x_r| |u_r| = |v| |w|$.  Then using the fact that
 \eqn\Fproperty{ \FF( |X|^2) = \FF \left( \frac{ |v|^2 |w|^2 }{|X|^2} \right)~,}
 one easily verifies that the two definitions agree up to a transformation of the sort \KKahlertransres:
\eqn\Ktransition{ \KK^{(-)} = \KK^{(+)} + \frac{8 L}{\sqrt{3}} y {\rm Re}( \log{v} + \log{w} )~.}
We have previously remarked that the geometry \gcnice\ is invariant under combined transformations of the form \xtransition, \Ktransition; in particular solutions of the Monge-Ampere equation are invariant under \Ktransition.  Therefore we have demonstrated that \yglobal, \xsevenglobal, and \Kglobal\ give a well defined parameterisation of the conifold, through \gcnice, everywhere except $x = u = v = 0$ and $x = u = w = 0$.  We have claimed that these points correspond to the brane locus.  To verify that claim, and complete the map between \gcnice\ and the conifold, we now show that $\KK$ has the correct singularity structure as we approach the brane locus, as dictated by \Mongenice.

Part of the challenge is that we must determine the appropriate ``near-brane'' coordinates $\eta,\lambda$, \nearbrane.  Note that, as we approach the brane locus, the change of variables $(v,w) \mapsto (\lambda(v,w), \eta(v,w))$ should have unit determinant.  Then the Monge-Ampere equation, \Mongenice, will be invariant under the change of variables, and the form \nearbrane\ that $\KK$ must take\foot{The near-brane behavior \nearbrane\ is appropriate for the examples considered in this paper.  However, for more general five-brane webs it appears that \nearbrane\ is too strong: the branes should indeed follow holomorphic profiles, but the tangential metric along the brane worldvolume need not be flat \MR.} in terms of $\eta,\lambda$, is consistent with \Mongenice.

We expect that the brane locus is described by $v w = y = 0$.  Let us zoom in towards a point on this locus where $y,w \to 0$, but $v$ remains finite:
\eqn\NSzoom{ y = \epsilon {\hat y}~, \qquad w = \epsilon \hat{w}~,}
with $\epsilon$ small, and ${\hat y},{\hat w},v \sim \OO(1)$.  Thus we are zooming in close to the $NS$-brane, while remaining far away from the $NS'$-brane.  (Since we are on the geometric side of the T-duality, there are not actually any branes present, but the sources in the equation of motion for $\KK$ are still present).  $({\hat w},{\hat y})$ parameterise a region of $\IR^3$ about the $NS$-source, which is located at the origin of this $\IR^3$.  Consistency of equation \yglobal\ for $y$ imposes the leading order scaling behaviors $|x|^2 \sim \OO(\epsilon)$, $|u|^2 \sim \OO(\epsilon)$.  We can set
\eqn\xpert{ |x|^2 = \epsilon \left( |x_0|^2 + \epsilon |x_1|^2 + \cdots \right)~, \qquad |u|^2 = \epsilon \left( |u_0|^2 + \epsilon |u_1|^2 + \cdots \right)~,}
and solve \yglobal\ perturbatively for the subleading corrections.  Before doing so, however, we observe that \NSzoom, \xpert\ imply
\eqn\Kzero{ \KK^{(\pm)} = \frac{3}{2} L^{2/3} |v|^{4/3} + \OO(\epsilon)~.}
It follows that the tangential coordinate, $\eta$, should be given by $\eta = \sqrt{3} L^{1/3} v^{2/3}$, up to possible $\OO(\epsilon)$ corrections---$\KK$ only needs to take the form \nearbrane\ in the limit $\epsilon \to 0$.  We can fix the normal coordinate, $\lambda$, by imposing that the Jacobian for the map $(v,w) \mapsto (\lambda,\eta)$ has unit determinant as $\epsilon \to 0$.  Doing so, we find that the near-brane variables should be
\eqn\etalambda{ \lambda = \frac{\sqrt{3}}{2L^{1/3} }v^{1/3} w + \OO(\epsilon^2)~, \qquad \eta =\sqrt{3} L^{1/3} v^{2/3} + \OO(\epsilon)~.}
Since $\lambda \sim \OO(\epsilon)$, we define $\lambda = \epsilon {\hat \lambda}$.  The inverse of \etalambda\ is
\eqn\vwetalambda{ {\hat w} = \frac{2 \sqrt{L}}{3^{1/4} \sqrt{\eta}} {\hat \lambda} + \OO(\epsilon)~, \qquad v = \frac{1}{3^{3/4} \sqrt{L}} \eta^{3/2} + \OO(\epsilon)~.}

We can now proceed to solve \coordtrans\ perturbatively for $|x|^2,|u|^2$.  We plug \NSzoom, \xpert, \vwetalambda\ into \yglobal, and find that the leading order solution is
\eqn\leadingx{ |x_{0}|^2 = \frac{2}{3} |\eta| \left[ {\hat y} + \sqrt{ {\hat y}^2 + | {\hat \lambda} |^2 } \ \right]~, \qquad |u_0|^2 =  \frac{2}{3} |\eta| \left[ - {\hat y} + \sqrt{ {\hat y}^2 + | {\hat \lambda} |^2 } \ \right]~.}
Subleading terms in $|x|^2,|u|^2$ will depend on the subleading terms in \vwetalambda, but \leadingx\ will be sufficient for our purposes.  Plugging \vwetalambda\ and \leadingx\ into \Kglobal\ and expanding yields
\eqn\Kzeroexpansion{ \KK^{(\pm)} = \frac{1}{2} |\eta|^2 + \frac{4 L}{\sqrt{3}} \left\{ \sqrt{ y^2 + |\lambda|^2} \mp y \log{ \left[ \frac{2 |\eta|}{3 c^2} \left( \pm y + \sqrt{y^2 + |\lambda|^2} \ \right) \right]} \right\} + \OO(\epsilon^2)~.}
In order for $\KK$ to be well defined, we must choose $\KK = \KK^+$ on the positive $y$-axis and $\KK = \KK^-$ on the negative $y$-axis.  When both $y,\lambda \neq 0$ we are free to choose either expression, since they are related by the gauge transformation \Ktransition.  In the neighborhood of $y=0$, however, there is a unique choice that extends $\KK$ to $y=0$ such that \Mongenice\ is satisfied: we take $\KK = \KK^+$ as $y \to 0_+$ and $\KK = \KK^-$ as $y \to 0_-$.  This may be conveniently expressed as
\eqn\Kexpansion{ \eqalign{ & \KK = \frac{1}{2} \eta {\bar\eta} + \KK_2(\lambda,{\bar\lambda},y) + {\tilde \KK}_{2}(y,\eta,{\bar\eta}) + \OO(\epsilon^2)~, \qquad {\rm with} \cr
& \KK_2(y,\lambda,{\bar\lambda}) = \frac{4 L}{\sqrt{3}}  \left\{ \sqrt{ y^2 + |\lambda|^2} - |y| \log{ \left[  |y| + \sqrt{y^2 + |\lambda|^2}  \right]} \right\} ~,}}
and where ${\tilde \KK}_2 \sim y \log(\eta {\bar\eta})$ makes no contribution to the equation of motion.

The singularity structure of $\KK_2$ is precisely that required by \Mongenice.  Taking $y$-derivatives carefully, we have
\eqn\dyKtwo{ \eqalign{ \d_y \KK_2 =&~ -\frac{4 L}{\sqrt{3}} \sgn(y) \log{ \left[ |y| + \sqrt{ y^2 + |\lambda|^2} \right]} \cr
\Rightarrow \quad \d_{y}^2 \KK_2 =&~ - \frac{4 L}{\sqrt{3}} \left[ \delta(y) \log{|\lambda|^2} + \frac{1}{\sqrt{ y^2 + |\lambda|^2} } \right]~,}}
while
\eqn\dLambdaKtwo{ \d_\lambda \delbar_{\bar\lambda} \KK_2 = \frac{L}{\sqrt{3} \sqrt{ y^2 + |\lambda|^2}}~.}
Therefore \Mongenice\ is satisfied, with ${\tilde Q} = \frac{8\pi}{\sqrt{3}} L$.  Furthermore, from \gAK\ we learn that the warp factor near the brane behaves as
\eqn\warpnearbrane{ e^{-3A} = \frac{2 L}{\sqrt{3} \sqrt{ y^2 + |\lambda|^2}}~,}
plus corrections that are finite in the limit $y,\lambda \to 0$.  By comparing this result with the standard form of the smeared CHS geometry \refs{\BanksRJ,\CallanAT,\GregoryTE}, we determine the scale $L$ in terms of the asymptotic radius ${\tilde R}_7$:
\eqn\Lfix{ \frac{4 L}{\sqrt{3}} = {\tilde R}_7~.}
This is the relation one might have naively expected from \coordtrans, since it gives ${\tilde x}^{7} = {\tilde R}_7 \phi_x$.  It also gives a standard result for the charge, ${\tilde Q} = 2\pi {\tilde R}_7$, in \Mongenice, corresponding to one $NS5$-brane smeared on the dual circle.

We stress that \Lfix\ is a nontrivial piece of data about the full asymptotically flat geometry describing the brane intersection, and its T-dual.  The scale $L$, introduced in \Rflatcon, \Fresult, describes the overall ``size'' of the conifold.  We have learned that in the full geometry, the size of the conifold throat is not arbitrary, but is determined by the asymptotic radius of the ${\tilde x}^7$ circle.

We can instead zoom in towards the $NS'$-brane:
\eqn\NSpzoom{ y = \epsilon' {\hat y}'~, \qquad v = \epsilon' {\hat v}'~,}
with $\epsilon'$ small and ${\hat y}', {\hat v}', w \sim \OO(1)$.  Since the configuration has a ${\bf Z}_2$ symmetry $v \leftrightarrow w$, the near-brane coordinates should take the form,
\eqn\nearNSprime{  \lambda = \frac{\sqrt{3}}{2L^{1/3} }w^{1/3} v + \OO({\epsilon'}^2)~, \qquad \eta =\sqrt{3} L^{1/3} w^{2/3} + \OO(\epsilon')~,}
and the rest of the analysis will follow.  The classical geometry is singular at the point $y = v = w = 0$, corresponding to the conifold singularity, and we will not consider it.

Computing the mixed derivatives $\KK_{yv}$ from \Kexpansion\ leads to terms that involve $\sgn(y)$.  This discontinuity at $y = 0$ can be eliminated in $\KK_{yv}^{\rm reg.}$ by choosing ${\tilde f}$ in \Bfinal\ appropriately, so that the circle fibre is smooth.  The discussion is identical to the Taub-NUT case, which is presented in Appendix C.

In summary, we have demonstrated that $\KK$ is a solution to the sourced Monge-Ampere equation, \Mongenice.  Thus, \gcnice\ (with \yglobal, \xsevenglobal, and \Kglobal) describes the singular conifold, while \NSfive\ gives the geometry of two intersecting $NS5$-branes smeared on a transverse circle.  In Appendix B we also show that the Killing spinors of \gcnice\ are consistent with the standard result for the Killing spinors of the conifold\foot{The Killing spinors of the conifold are charged under the angular coordinate $\psi$.  Since the T-duality direction ${\tilde x}^7$, \xsevenglobal, has a component along $\psi$, one might conclude that this T-duality should break supersymmetry.  The results of Appendix B explicitly show that this is not the case.  Heuristically speaking, we avoid this conclusion because, via a coordinate reparameterisation \inducedshift\ and compensating gauge transformation \KKahlertransres\ on $\KK$, we can make ${\tilde x}^7$ independent of $\psi$ everywhere, except on a set of measure zero.}.  We will discuss the intersecting brane geometry, \NSfive, further in section 3.5, but now we turn to Legendre transformations.

\subsec{The Legendre transformation}

We have already remarked that the coordinate $y$ and the potential $\KK$ can be obtained via Legendre transformation of the K\"ahler potential $\FF$ of the conifold, where the dual variable is $\log{|x|^2}$ (or $\log{|u^2|}$).  Legendre transformations have arisen in similar contexts before \refs{\LindstromRT,\HitchinEA,\RocekPS,\KiritsisWD,\AndreasHH}, both from sigma model and spacetime points of view.  The Legendre transformation between symplectic and K\"ahler coordinate systems has also been discussed from the mathematical point of view in \Abreu\ (see also \BalasubramanianDV)\foot{In those cases the transformation was performed on all coordinates paired to $U(1)$ isometry directions; here $\KK$ is defined by Legendre transformation of the K\"ahler potential on a single coordinate.  It is thus a hybrid between the K\"ahler potential and symplectic potential of \Abreu.}.

We will now show that any K\"ahler metric with a $U(1)$ isometry can be brought to the form \gcnice\ by such a Legendre transformation, and we will find explicit expressions for the components $\KK_{mn}$, $m,n = y,a,\abar$, in terms of $\FF$.  Working on the patch $x \neq 0$, we will also show that Ricci flatness of the K\"ahler metric implies the source free Monge-Ampere equation, \Monge, for the potential $\KK$.    (This last result was given in the original work of \LindstromRT; we provide a brief derivation here for completeness).

Let ${\cal M}$ be a complex-dimension $n+1$ K\"ahler manifold with a $U(1)$ isometry.  Locally we may choose coordinates $\zeta^{\alpha} = (x,z^a)$, $a = 1,\ldots,n$, such that the isometry direction is identified with the phase of $x$.  Let $x = c e^{\xi/2} e^{i \phi_x}$, so that $\xi = \log{(|x|^2/c^2)}$.  $c$ is a real constant inserted to get dimensions correct.  Then our assumption is that the K\"ahler potential on ${\cal M}$ is independent of $\phi_x$:
\eqn\Fded{ \FF = \FF(\xi, z^a,\zbar^\abar)~.}
We denote the components of the K\"ahler metric by $g_{\alpha{\bar\beta}} = \d_\alpha \delbar_{\bar\beta} \FF$.  Changing variables from $(x,\xbar,z^a,\zbar^\abar)$ to $(\xi,\phi_x,z^a,\zbar^a)$ brings the K\"ahler metric to the form
\eqn\xiphimet{ \eqalign{ ds^2 =&~ 2 \left[ g_{x\xbar} dx d\xbar + g_{x \bbar} dx d\zbar^\bbar + g_{a \xbar} dz^a d\xbar + g_{a\bbar} dz^a d\zbar^\bbar \right] \cr
=&~ c^2 e^{\xi} g_{x\xbar} \left( \frac{1}{2} d\xi^2 + 2 d\phi_{x}^2 \right) + c e^{\xi/2} e^{i \phi_x} g_{x\bbar} \left( d\xi + 2 i d\phi_x \right) dx d\zbar^\bbar + \cr
& \qquad \qquad \qquad \qquad + c e^{\xi/2} e^{-i\phi_x} g_{a\xbar} \left( d\xi - 2 i d\phi_x \right) dz^a d\xbar + 2 g_{a\bbar} dz^a d\zbar^\bbar~.}}

Next consider a change of variables
\eqn\KahlertoBrane{ (\xi, \phi_x, z^a, \zbar^\abar) \mapsto (y, {\tilde x}^7, z^a, \zbar^\bbar)~,}
assuming
\eqn\xiphitoyx{ \phi_x = \frac{1}{c} {\tilde x}^7~, \qquad \xi = \xi(y,z^a,\zbar^\bbar)~.}
After considerable but straightforward algebra, the metric may be expressed as
\eqn\Branemetone{ \eqalign{ ds^2 =&~ \frac{c^2}{2} e^\xi g_{x\xbar} (\d_y \xi)^2 dy^2 + 2 \left( g_{a\bbar} - \frac{ g_{a\xbar} g_{x\bbar}}{g_{x\xbar}} \right) dz^a d\zbar^\bbar + \cr
& \quad + 2 g_{x\xbar} e^\xi \left[ d{\tilde x}^7 - \frac{i}{2 g_{x\xbar}} e^{-\xi/2} \left( e^{-i\phi_x} g_{a\xbar} dz^a - e^{i\phi_x} g_{x\bbar} d\zbar^\bbar \right) \right]^2 + \cr
& \quad + c e^{\xi/2} \d_y \xi g_{x\xbar} \left[ \left( c e^{\xi/2} \d_a \xi + e^{-i\phi_x} \frac{g_{a\xbar}}{g_{x\xbar}} \right) dz^a + \left( c e^{\xi/2} \delbar_\bbar \xi + e^{i\phi_x} \frac{g_{x\bbar}}{g_{x\xbar}} \right) d\zbar^\bbar \right] dy + \cr
& \quad + \frac{1}{2} g_{x\xbar} \left[  \left( c e^{\xi/2} \d_a \xi + e^{-i\phi_x} \frac{g_{a\xbar}}{g_{x\xbar}} \right) dz^a + \left( c e^{\xi/2} \delbar_\bbar \xi + e^{i\phi_x} \frac{g_{x\bbar}}{g_{x\xbar}} \right) d\zbar^\bbar \right]^2 ~.}}
We also remind the reader of some standard partial derivative relations that follow from the change of variables \KahlertoBrane\ and its inverse:
\eqn\partialrelations{  \d_y \xi = \frac{1}{\d_\xi y}~, \quad \left( \d_a \xi \right)_{(y)} = - \left( \d_y \xi \right)_{(z^a)} \left( \d_a y \right)_{(\xi)}~, \quad \left( \del_\abar \xi \right)_{(y)} = - \left( \d_y \xi \right)_{(\zbar^\abar)} \left( \delbar_\abar y \right)_{(\xi)}~.}
The notation $( \cdot )_{(k)}$ is to emphasize that the derivative is being taken while the quantity $k$ is held fixed.

Expressions \Branemetone, \partialrelations\ hold for any $\xi = \xi(y,z^a,\zbar^\abar)$, but now let us consider the special case where the function $\xi$ is determined by inverting the expression
\eqn\xiextremize{ c y = \d_\xi \FF~.}
A similar change of variables was considered in \ChenDU.  Since $\FF$ only depends on $x$, $\xbar$ through $\xi$, we have $g_{x\xbar} = \d_x \delbar_\xbar \FF = \frac{1}{c^2} e^{-\xi} \d_{\xi}^2 \FF$.  Similarly, $g_{a \xbar} = \frac{1}{c} e^{-\xi/2} e^{i\phi_x} \d_a \d_\xi \FF$.  Using \xiextremize\ and \partialrelations\ we get
\eqn\gxxgax{ g_{x\xbar} = \frac{e^{-\xi}}{c \d_y \xi}~, \qquad g_{a\xbar} = - e^{-\xi/2} e^{i\phi_x} \frac{\d_a \xi}{\d_y \xi}~, \qquad g_{x\abar} = - e^{-\xi/2} e^{-i \phi_x} \frac{ \delbar_\abar \xi}{ \d_y \xi}~.}
With these relations the last two lines of \Branemetone\ vanish and the metric becomes
\eqn\Branemettwo{  \eqalign{ ds^2 =&~ \frac{c}{2} \d_y \xi dy^2 + 2 \left( g_{a\bbar} - c \frac{ \d_a \xi \delbar_\bbar \xi}{\d_y \xi} \right) dz^a d\zbar^\abar + \cr
& \qquad \qquad \qquad \qquad \qquad + \frac{2}{c \d_y \xi} \left[ d {\tilde x}^7 + \frac{ic }{2} \left( \d_a \xi dz^a - \delbar_\abar \xi d\zbar^\abar \right) \right]^2~.}}

Now introduce the potential $\KK$ as the Legendre transform of $\FF$:
\eqn\KLegendredef{ \KK(y,z^a,\zbar^\abar) =  \left\{ \FF(\xi,z^a,\zbar^\abar) - c y \xi \right\}_{{\rm max}~\xi} ~.}
Let us compute the various partials of $\KK$ from this definition, keeping in mind that $\FF$ depends on $y$ implicitly through $\xi$, and on $z^a,\zbar^\abar$ both explicitly and implicitly.  We find
\eqn\Kmnder{ \KK_{yy} = - c \d_y \xi~, \quad \KK_{ya} = -c \d_a \xi~, \quad \KK_{y\abar} = -c \delbar_\abar \xi~, \quad \KK_{a\bbar} = g_{a\bbar} - c \frac{ \d_a \xi \delbar_\bbar \xi}{\d_y \xi}~,}
and using these results, the metric \Branemettwo\ is given by
\eqn\gcniceagain{ ds^2 = - \frac{1}{2} \KK_{yy} dy^2 + 2 \KK_{a\bbar} dz d\zbar^\bbar - \frac{2}{\KK_{yy}} \left[ d{\tilde x}^7 - \frac{i}{2} \left( \KK_{ya} dz^a - \KK_{y\abar} d\zbar^\abar \right) \right]^2~.}
Now, as we are about to show, if the K\"ahler metric is Ricci flat, then $\KK$ constructed in this fashion satisfies the source-free Monge-Ampere equation \Monge.  In this case $-\frac{1}{2} \KK_{yy} = e^{-3A}$, $\KK_{ya} = \KK_{ya}^{\rm reg.}$, and \gcniceagain\ reproduces \gcnice.

Let us demonstrate that Ricci flatness of the metric $g_{\alpha{\bar\beta}}$ implies the Monge-Ampere equation, \Monge, and vice versa.  First we note that \gxxgax\ may also be written as
\eqn\gxxgaxtwo{ g_{x\xbar} = \frac{1}{c} e^{-\xi} \d_\xi y~, \qquad g_{x\abar} = e^{-\xi/2} e^{-i \phi_x} \delbar_\abar y~, \qquad g_{a\xbar} = e^{-\xi/2} e^{i\phi_x} \d_a y~.}
Then consider the determinant of the K\"ahler metric:
\eqn\detKahler{ \eqalign{ g =&~ \left| \matrix{  \frac{1}{c} e^{-\xi} \d_\xi y & e^{-\xi/2} e^{-i \phi_x} \delbar_\bbar y \cr  e^{-\xi/2} e^{i\phi_x} \d_a y & g_{a\bbar} } \right| = \frac{e^{-\xi}}{c^2} \left| \matrix{ c \d_\xi y & c \delbar_\bbar y \cr c \d_a y & g_{a\bbar} } \right|  \cr
=&~ \frac{e^{-\xi}}{c} \d_\xi y \ \det{\left( g_{a\bbar} - c \frac{ \d_a y \delbar_\bbar y}{\d_\xi y} \right)}~.}}
Making use of \partialrelations, \Kmnder\ we get
\eqn\detKahlerthree{ g = - \frac{ e^{-\xi} }{\KK_{yy}} \det_n{(\d \delbar \KK)}~.}
The Ricci tensor on a K\"ahler manifold is $R_{\alpha{\bar\beta}} = \d_\alpha \delbar_{\bar\beta} \log{g}$, and therefore, on the patch $x \neq 0$, it is clear that the Monge-Ampere equation \Monge, (with $\KK$ independent of ${\tilde x}^7$), implies Ricci flatness.

A little more work is required to go the other way.  The general conclusion we may draw from $R_{\alpha{\bar\beta}} = 0$ is that $g = |f(x,z^a)|^2$, for some holomorphic function $f$.  However, using the fact that $\log{g}$ only depends on $x,\xbar$ through $\xi$, we can refine this condition to
\eqn\Ricciimplies{ R_{\alpha{\bar\beta}} = 0 \quad \Rightarrow \quad g = |x|^{2B} |f(z^a)|^2~,}
where $B$ is an arbitrary real constant.  But now, getting \Monge\ out of \Ricciimplies\ is just a matter of choosing the right coordinates.  Consider a holomorphic reparameterisation $z^a \to {z'}^{a'}$ given by $z^a = z^a({z'}^{a'})$, combined with $x \to x'$ given by $x/c = (x'/c)^\lambda$.  This corresponds to a rescaling $(\xi,\phi_x) = \lambda (\xi' , \phi_x')$, which induces a transformation $(y,{\tilde x}^7) = (\lambda^{-1} y', \lambda {\tilde x}^{7'})$ on the dual coordinates.  Under this transformation,
\eqn\reparameterise{ g = \lambda^2 |x|^{2(\lambda -1)} |J(z)|^2 g' = \lambda^2 |x|^{2(\lambda -2)} |J(z)|^2 \left( - \frac{ \det_{n}'(\d \delbar \KK) }{ \KK_{y' y'} } \right)~,}
where $J(z)$ is the Jacobian for $z^a \to {z'}^{a'}$.  Now choose the new coordinates such that\foot{If $B = -2$ then the correct change of coordinates is $x = c e^{x'/c}$.  The isometry in $\phi_x$ maps to a shift isometry in ${\rm Im}(x')$.  In this case it is natural to identify $\xi' \equiv \frac{2}{c} {\rm Re}(x') = \xi$, implying $y' = y$.}
\eqn\coordfix{ \lambda = B + 2~, \qquad J(z) = \frac{2\sqrt{2}}{\lambda^2} f(z)~.}
Plugging \reparameterise\ into \Ricciimplies\ and rearranging leads to the generalisation of \Monge\ for arbitrary dimension:
\eqn\Mongedimn{ \d_{y'}^2 \KK + 8 \det_{n}'( \del \delbar \KK) = 0~.}

The fact that a particular choice of coordinate system $(x,z^a)$ is necessary in order to derive \Mongedimn\ from Ricci flatness is not surprising since \Mongedimn\ is not covariant under changes of coordinates.  Expressed in terms of the K\"ahler metric determinant, the choice of coordinates should be such that $g = e^{-\xi}/8 = c^2/(8|x|^2)$.  One can check that this holds for the case of the conifold, defined by K\"ahler potential \Fresult, using the relation $c = \frac{4 L}{\sqrt{3}}$.

In summary, given a K\"ahler manifold with K\"ahler potential $\FF$, a $U(1)$ isometry, and a local coordinate system $(x,z^a)$, such that the phase of $x$ is associated with the $U(1)$ direction, we can make a change of coordinates $(x,\xbar) \to (y,{\tilde x}^7)$, where ${\tilde x}^7 \propto \phi_x$, $y \propto \d_{\log{|x|^2}} \FF$, such that the metric is described by \gcniceagain, where $\KK$ is the Legendre transform of $\FF$ with respect to the dual pair $(y,\log{|x|^2})$.  Furthermore, if the manifold is Ricci flat, then we can choose the coordinate system $(x,z^a)$ such that $\KK$ satisfies the source-free Monge-Ampere equation \Mongedimn.  We emphasise that these are all local statements.  If the K\"ahler manifold is T-dual to a web of $NS5$-branes, described by \NSfive, one expects that the brane locus should be described by a surface where this local patch breaks down---for exampe, the surface $x = 0$ where the phase is no longer defined---so that there is a possibility for source terms in \Mongedimn.  It would be interesting to study these global issues from a general point of view, but we will not pursue them here.

\subsec{The intersecting brane geometry}

In this section we collect the results from previous sections and write down the supergravity background corresponding the $NS$-$NS'$ pair, smeared on the $x^7$ circle.  Recall that the supergravity solution is completely determined by a single function $\KK(y,z^a,\zbar^\abar)$, $(z^1,z^2) \equiv (v,w)$.  In string frame we have
\eqn\NSfivetwo{ \eqalign{ & ds^2 = -dt^2 + d{\bf x}_{3}^2 + 2 \KK_{a\bbar} dz^a d\zbar^{\bbar} + e^{-3A} ( dy^2 + R_{7}^2 d\phi_{7}^2 )~, \cr
& e^{\varphi - \varphi_0} = e^{-3A/2}~, \cr
& H_3 = - i R_7 \left( \d_a e^{-3A} dz^a - \delbar_{\abar} e^{-3A} d\zbar^\abar \right)dy d\phi_7 + i R_7 \d_y \KK_{a\bbar} dz^a d\zbar^\bbar d\phi_7~,}}
with
\eqn\warpagain{ e^{-3A} = 4 \det{(\d \delbar \KK)}~.}
In order to define $\KK$, we must introduce an auxiliary function $|X|^2 = |X|^2(y,|v|,|w|)$, determined by the solution to
\eqn\Xaux{ |y| = \left( \frac{3 R_7}{2 \ell_{s}^2} \right)^{1/3} \frac{\left( |X|^4 - |v|^2 |w|^2 \right) }{ 2 |X|^{4/3} \left( |X|^2 + |v|^2 \right)^{1/3} \left( |X|^2 + |w|^2 \right)^{1/3} }~, \qquad |X|^2 \geq |v| |w|~.}
Then the brane potential is
\eqn\branepot{\KK(y,|v|,|w|) =  \left( \frac{9 \ell_{s}^2}{4 R_7} \right)^{2/3} \frac{ (|X|^2 + |v|^2)^{2/3} (|X|^2 + |w|^2)^{2/3} }{2 |X|^{4/3} } - \frac{\ell_{s}^2 |y|}{R_7} \log{\left( \frac{R_{7}^2 |X|^2}{\ell_{s}^4} \right)}~.}
In translating from the results of sections 3.2, 3.3, we have used $\frac{4 L}{\sqrt{3}} = c = {\tilde R}_7 = \ell_{s}^2/R_7$.

Our analysis in sections 3.2, 3.3, guarantees that this potential solves the appropriate sourced nonlinear PDE, \Mongenice.  In the language of section 3.3, we have chosen to work on the $|x| >0$ patch for $y \geq 0$ and the $|u| >0$ patch for $y \leq 0$.  This choice allows for a more compact presentation of the solution.

The coordinates $(y,v,w)$ are the natural ones from the intersecting brane point of view.  In these coordinates the $NS$-brane lies along $w = y = 0$, while the $NS'$ sits at $y = v= 0$, and $y$ describes the overall orthogonal direction to the stack (as is clear from the initial ansatz \Dfive).  However, they are somewhat inconvenient since expressions for metric components, the dilaton, and $H_3$ must be given in terms of the auxiliary function $|X|^2(y,|v|,|w|)$ which is, in general, the root of a sextic polynomial\foot{On special surfaces such as $v = 0$ or $w = 0$, the degree is reduced and explicit expressions in terms of $(y,|v|,|w|)$ may be given.}.  This situation can be ameliorated by introducing a different coordinate system.

We introduce conifold-like coordinates $(\rho,\theta_1,\theta_2,\phi_v,\phi_w)$, defined by
\eqn\concoords{ \eqalign{  v =&~ \frac{2 \sqrt{R_7}}{3 \ell_{s}} \rho^{3/2} \cos{ \frac{\theta_1}{2} } \sin{ \frac{\theta_2}{2}} e^{i \phi_v}~, \cr
w =&~ \frac{2 \sqrt{R_7}}{3 \ell_{s}} \rho^{3/2} \sin{ \frac{\theta_1}{2} } \cos{ \frac{\theta_2}{2} } e^{i \phi_w}~, \cr
 y =&~ \frac{R_7}{6\ell_{s}^2} \rho^2 (\cos{\theta_1} + \cos{\theta_2})~.}}
The definitions for $v,w$ agree with those in \changeofvar, using \rhodef\ and $L = \frac{\sqrt{3} \ell_{s}^2}{4 R_7}$, and where $\phi_v = \frac{1}{2}(\psi + \phi_1 - \phi_2) - \pi$ and $\phi_w = \frac{1}{2}(\psi - \phi_1 + \phi_2)$.  It is more convenient to use $\phi_v, \phi_w$ since, as we saw above, the $x^7$ circle is identified with $\phi_x$ or $\phi_u$ under T-duality (and not $\psi$).  The definition for $y$ follows from \Xaux\ and \changeofvar, using $X = x$ for $y \geq 0$ and $X = u$ for $y \leq 0$.  The coordinates $\theta_1,\theta_2$ take values in $[0,\pi]$.  We see that the $NS$-brane is located at $\theta_1 = 0$, $\theta_2 = \pi$, while the $NS'$-brane is located at $\theta_1 = \pi$, $\theta_2 = 0$.  The positive $y$-axis corresponds to $\theta_1 = \theta_2 = 0$, while the negative $y$-axis corresponds to $\theta_1 = \theta_2 = \pi$.  The surface $y = 0$ corresponds to the line $\theta_1 + \theta_2 = \pi$, connecting the $NS$ and $NS'$ at opposite corners of the $\theta_1,\theta_2$ plane $[0,\pi]\times [0,\pi]$. The brane intersection at $v = w = y = 0$ corresponds to $\rho = 0$.

One can similarly write down an expression for $\KK$ in terms of $(\rho, \theta_1,\theta_2)$, but this is somewhat deceptive.  The equation of motion, \Mongenice, satisfied by $\KK$ is not covariant; transforming it to the new coordinate system \concoords\ leads to a hideously complicated equation.  It is better to find expressions for $\KK_{mn}$, $m = y,a,\abar$, in terms of $(|X|,|v|,|w|)$ directly from \Xaux, \branepot, (such as in equations \Kwwbarvvbar, \etc.),  and then plug in \concoords.  Doing this, and taking into account the usual factors from the Jacobian for the change of variables, we find the following explicit form of the metric and dilaton:
\eqn\metdilradial{ \eqalign{ & ds^2 = -dt^2 + d{\bf x}_{3}^2 + d\rho^2 + \rho^2 d\Omega_{4}^2 + e^{-3A} R_{7}^2 d\phi_{7}^2~, \cr
& e^{2(\varphi - \varphi_0)} = e^{-3A} = \frac{18 \ell_{s}^4}{R_{7}^2 \rho^2 \left[ 6 (1+ c_1 c_2) - (c_1+c_2)^2 \right] }~,}}
where $c_i \equiv \cos{\theta_i}~$, $s_i \equiv \sin{\theta_i}$.  The compact four-manifold is parameterised by $(\theta_1,\theta_2,\phi_v,\phi_w)$ and has metric
\eqn\fourmetric{ \eqalign{ d\Omega_{4}^2 = \frac{1}{6} (d\theta_{1}^2 + d\theta_{2}^2 ) & + \frac{2\left[7 + c_1 c_2 - 3(c_1-c_2) \right]}{3\left[6(1+c_1c_2)-(c_1+c_2)^2 \right]} \cos^2{\frac{\theta_1}{2}} \sin^2{\frac{\theta_2}{2}} d\phi_{v}^2 + \cr
& + \frac{ 2\left[7 + c_1 c_2 + 3(c_1-c_2) \right]}{3 \left[6(1+c_1c_2)-(c_1+c_2)^2\right] } \sin^2{\frac{\theta_1}{2}} \cos^2{\frac{\theta_2}{2}} d\phi_{w}^2 + \cr
& + \frac{ s_{1}^2 s_{2}^2}{3\left[6(1+c_1c_2)-(c_1+c_2)^2\right]} d\phi_v d\phi_w~.}}

The Neveu-Schwarz three-form flux is
\eqn\Hthreeexplicit{ \eqalign{ H_3 =&~ 4 \ell_{s}^2 \left\{ \frac{6 + c_{1}^2 + c_{2}^2 - c_1c_2 }{\left[ 6(1+c_1 c_2) - (c_1 + c_2)^2 \right]^2} \left( \sin^2{\frac{\theta_2}{2}} d\phi_v + \cos^2{\frac{\theta_2}{2}} d\phi_w \right) + \right. \cr
& \qquad ~ \left. + \frac{3 (2 c_2 - c_1) }{\left[ 6(1+c_1 c_2) - (c_1 + c_2)^2 \right]^2}  \left( \sin^2{\frac{\theta_2}{2}} d\phi_v - \cos^2{\frac{\theta_2}{2}} d\phi_w \right) \right\} s_1 d\theta_1 d\phi_7 + \cr
&  +  (\theta_1,\phi_v,\phi_w) \leftrightarrow (\theta_2, \phi_w,\phi_v)~.}}
We can check that this corresponds to the right charges by computing $- \int_{\Sigma_3} H_3$ over three-surfaces enclosing each brane.  A nice choice for the $\Sigma_3$'s is to set $\theta_1 = \theta_2 \equiv \theta$.  In this case
\eqn\Hthreespecial{ H_3 \displaystyle\bigg|_{\theta_1 = \theta_2}  = -2 \ell_{s}^2 \frac{(3 - \cos^2{\theta})}{(3 + \cos^2{\theta})^2} \sin{\theta} d\theta (d\phi_v + d\phi_w) d\phi_7~.}
We integrate over $\theta \in [0,\pi]$, $\phi_7$, and $\phi_v$ $(\phi_w)$ to enclose the $NS$-brane ($NS'$-brane), finding
\eqn\intflux{ -\frac{1}{2} \int_{\Sigma_{3}^{NS}} H_3 = -\frac{1}{2} \int_{\Sigma_{3}^{NS'}} H_3  = 2\pi^2 \ell_{s}^2~.}
This is the standard result for the charge of a single $NS5$-brane.

While we leave a thorough study of this geometry for future work, we can not resist pointing out some interesting features.
\item{(1)}  The dilaton \metdilradial\ blows up at $(\theta_1, \theta_2) = (0,\pi)$ and $(\pi,0)$, as well as $\rho = 0$.  These are precisely the locations of the $NS$ and $NS'$ consistent with the behavior of the dilaton for a single $NS5$-brane.
\item{(2)}  It appears that the $x^7$ circle never gets larger than string scale (unless $\rho \ll \ell_s)$.  However, this is due the fact that \metdilradial\ {\it only describes the near horizon geometry, or throat region, of the $NS$-$NS'$ system}.  An asymptotically flat solution, describing the complete brane intersection, would have $e^{\varphi - \varphi_0} \to 1$ as $\rho \to \infty$, and would asymptote to this solution for $\rho R_7 \ll \ell_{s}^2$ (or $\rho \ll {\tilde R}_7$).  A related fact we have already mentioned is that the conifold is T-dual to the throat region of the brane geometry, and not the full asymptotically flat geometry.  Though we only know the near horizon geometry, we are able to nonetheless identify the parameter $R_7$ as the asymptotic radius of the circle in the full geometry.  We did this section 3.3 by determining the charge of the source and matching to standard results. 
\item{(3)}  Our analysis in section 3.3 shows that this geometry reduces to the appropriate form---the smeared CHS geometry---in the limit we approach one brane, remaining far away from the other.  It is amusing to consider the behavior of the dilaton along the $y$-axis, where $(\theta_1,\theta_2) = (0,0), (\pi,\pi)$ for positive or negative values of $y$, respectively.  From \concoords\ and \metdilradial\ one finds
\eqn\dilatonyaxis{ e^{2(\varphi - \varphi_0)} \displaystyle\bigg|_{y\rm{-axis}} = \frac{3 \ell_{s}^2}{4 R_7 |y|}~.}
Compare this with the (near horizon) result for the case of $n$ (coincident) {\it parallel} $NS5$-branes, $e^{2(\varphi - \varphi_0)} = n \ell_{s}^2/(2 R_7 |y|)$.  The $NS$-$NS'$ system would appear to have an $n_{\rm eff} = 3/2$.  Of course, if we enclose either brane by a three-surface and compute $\int_{\Sigma_3} H_3$, we measure the charge of a single $NS5$-brane, $n=1$, as we saw in \intflux.  The point we are making here though, is that the solution is truly nonlinear---the warp function is in no sense the sum of harmonic functions for each brane.
\item{(4)}  Consider the four-manifold described by \fourmetric.  Viewed as a torus fibration over the $\theta_1,\theta_2$-plane, $[0,\pi] \times [0, \pi]$, we can observe the following facts.  The $\phi_v$ circle degenerates on the boundaries $\theta_1 = \pi$, $\theta_2 = 0$, while the $\phi_w$ circle degenerates on the boundaries $\theta_1 = 0$, $\theta_2 = \pi$.  Both circles degenerate at the corners $(\theta_1,\theta_2) = (0,0)$ and $(\pi,\pi)$, so these are smooth points in the geometry.  On the other hand, only one circle degenerates at the corners $(\theta_1, \theta_2) = (0,\pi)$ and $(\pi,0)$.  One can check that the curvature blows up at these points, consistent with the presence of sources there.  Though the curvature blows up, it is integrable.  One can compute the Euler character directly, finding $\chi = 2$.  The volume of the space can also be computed exactly.

\noindent
This geometry clearly deserves further study.  There are many possible avenues of investigation: one could study the worldsheet theory in this background, supersymmetric probe branes, or compactify it to five dimensions.  The singularity at $\rho = 0$ can be resolved by separating the $NS5$-branes in the $y$-direction or deforming the equation for the brane locus $v w = 0 \to vw = \epsilon$.  These configurations are T-dual to the resolved and deformed conifold, respectively, and the corresponding brane geometries will be constructed in a follow-up to this work \MR.  It also seems likely that \metdilradial\ has a holographic interpretation, and the natural candidate theory is the four-dimensional little string theory living on the intersection of the $NS$ and $NS'$.

\newsec{Discussion}

The original motivation for this paper was to understand the geometric dual of the $NS5$-brane systems considered in \refs{\GiveonUR,\KutasovKB}, extending the work of \BuicanIS. A necessary preliminary step in this programme is to understand the precise duality relation between the conifold and a pair of intersecting $NS5$-branes, the task we have undertaken in this paper. There are two main results in this paper:
\item{(1)} We have constructed the first known example of a supergravity background for a pair of $NS5$-branes intersecting on $\IR^{1,3}$ and localised in all but one direction.
\item{(2)} We have constructed the precise T-duality relation (at the level of supergravity) between a pair of intersecting $NS5$-branes and the conifold.

\noindent Some comments on these results are in order.

Firstly, most known examples of intersecting brane supergravity solutions are restricted to the regime of $D$-branes. Even then, they are smeared in a number of directions and do not typically involve all three types of transversal coordinates (DD, DN and ND). The solution we have constructed here is localised in all but one direction, depends on all three categories of transverse coordinates and describes intersecting $NS5$-branes.

Secondly, the T-duality relation between the conifold and intersecting $NS5$-branes is more subtle, and in some aspects different, than is described in \refs{\BershadskySP,\DasguptaSU,\UrangaVF} in a number of ways. The conifold is a K\"ahler manifold and is most naturally expressed in a set of K\"ahler coordinates viz \xuvw-\kahmetone. If one wishes to relate the conifold to an intersecting $NS5$-brane brackground, at the level of supergravity solutions, there are two steps.
\item{(1)} Apply the Buscher T-duality rules along the appropriate $U(1)$ isometry;
 \item{(2)} Show the background one ends up with is describing a pair of $NS5$-branes---that is, it solves the supergravity equations of motion with the appropriate source terms and exhibits the correct symmetries.

These two tasks are made easier using the detailed analysis of \LuninTF, who showed the requisite properties a supergravity background describing a web of intersecting membranes must possess. These properties are most elegantly described in a certain coordinate system: a hybrid of symplectic coordinates $(y,x^7)$ and K\"ahler coordinates $(v,w)$, in which the physical properties of the brane system become manifest. For example, a consequence of supersymmetry is the statement that the five-brane locus is described by a holomorphic curve in the $v$-$w$ plane, while the singularity structure and bosonic symmetries are manifest in terms of the orthogonal coordinates $y^i = (y,x^7)$. Finally, \LuninTF\ prove the properties of the supergravity background described in this coordinate system match what one would expect from a probe analysis of the same system.

What we have described in this paper is how to correctly accomplish the above two steps.  Suppose we wish to T-dualise from the conifold to the intersecting $NS5$-brane system.  Firstly, the T-duality direction is identified as ${\tilde x}^7$ in \xsevenglobal.  Note that it is a nontrivial circle fibration over the base space described as $\IR_y \times \IC_v \times \IC_w$ with the brane locus cut out.  One may view ${\tilde x}^7$ as the momentum coordinate of a symplectic pair of coordinates.  We then obtain the corresponding position coordinate, $y$, along with the symplectic-K\"ahler potential $\KK$, via Legendre transform, and express the conifold metric in the form \gcnice, following the discussion in section 3.4.  Then we apply the usual Buscher T-duality along ${\tilde x}^7$ to arrive at the intersecting brane metric \NSfive.  After demonstrating that $\KK$ solves \Mongenice\ with the appropriate singularity structure, the results of \LuninTF\  guarantee that the background describes a pair of intersecting $NS5$-branes.

Thirdly, we have also clarified a number of confusing issues in the literature. These include:
\item{(a)} The T-dual of the conifold is not asymptotically flat. Consequently, the T-dual of the conifold is not the complete background for a pair of intersecting $NS5$-branes; rather it is their near horizon limit.  This point has also been recently discussed in \MaldacenaMW.
\item{(b)} The correct $U(1)$ to T-dualise along is not the conifold coordinate $\psi$ as was proposed in \DasguptaSU; rather it is a nontrivial circle fibration parameterised in patches by linear combinations of $\psi,\phi_1,\phi_2$, as defined in \xsevenglobal.  This gives an explicit coordinatisation of the T-duality proposed in \UrangaVF.
\item{(c)} The authors of \DasguptaSU\ end up with a geometry that is a circle fibration over $\IR^2\times \IR^2$, and not the $S^2\times S^2$ topology that is expected from the conifold.  Our analysis demonstrates that this is a consequence of their smearing on the relatively transverse coordinates $v,w$, as well as the overall transverse coordinate ${\tilde x}^7$.  A result of our analysis is that the application of Buscher's T-duality rules maps the conifold geometry into the brane geometry localised in the relatively transverse directions, and vice versa.

\noindent Finally, there remain a number of directions to be pursued for future work.
\item{(1)} What is the geometry that is dual to the complete $NS5$-brane background? It is presumably asymptotically flat, with its near horizon limit being the conifold. A simple analogy is Taub-NUT versus Eguchi-Hanson. The former is asymptotically flat and is T-dual to a parallel stack of $NS5$-branes. The latter is not asymptotically flat, and may be thought of as the T-dual of the near horizon limit of the $NS5$-branes. We have found the analogue of the Eguchi-Hanson space (conifold). What is the analogue of Taub-NUT?
\item{(2)} Our technique is generalisable to K\"ahler manifolds with a U(1) isometry. That is, one can perform a Legendre transformation followed by a T-duality to construct an intersecting brane solution. In a sequel to this paper \MR\ we do this for the resolved and deformed conifolds, showing how moduli between the two backgrounds map. It would be interesting to do this for other such spaces, for example the generalised conifold, and in particular the suspended pinch point (SPP) singularity\foot{ The SPP singularity may be thought of as a partial resolution of a $\Z_2\times\Z_2$ singularity \MorrisonCS\ discussed in \refs{\BuicanIS}.}, as discussed in the context of metastable supersymmetry breaking by \BuicanIS.
\item{(3)} What is the field theory dual of our near horizon background \Ksingular? It presumably has a connection with little string theory, and the symmetries and properties of our geometry may shed some light on the definition of this mysterious theory.

\bigskip
\noindent{\bf Acknowledgements:} It is a pleasure to thank A. Giveon, N. Halmagyi, and O. Lunin for helpful discussions.
JM is supported by an EPSRC Postdoctoral Fellowship EP/G051054/1.  AR acknowledges support from DOE grant DE-FG02-96ER50959.

\appendix{A}{ Determining the $B$-field of the brane web}
\seclab\appbfield

In this section we extract $B_2$ by studying the Hodge dual of $H_7$ in \NSfive. The nonzero components of $H_7$ are
\eqn\Hseven{ (H_7)_{\mu\nu\rho\sigma mnp} = - \frac{7!}{3!  4!} \epsilon_{\mu\nu\rho\sigma} (G_3)_{mnp}~, \qquad {\rm where} \quad  G_3 = i e^{-3A} d (e^{3A} \KK_{a \bbar} dz d\zbar^\bbar)~,}
and $\epsilon_{\mu\nu\rho\sigma}$ is the Levi-Civita tensor in (flat) $\IR^{1,3}$. We use $\mu,\nu$ for spacetime indices and $m,n$ for ``internal'' indices.  Taking the Poincare dual, we have the nonzero components of $H_3$ given by
\eqn\Hthree{ \eqalign{ (H_3)_{mnp} & = \frac{1}{3!}  {\epsilon}_{mnp}^{\phantom{mnp}qrs} (G_3)_{qrs}}~.}
The six dimensional Poincare dual is taken with respect to the metric
\eqn\sixmetric{ g_{mn} dy^m dy^n = 2 \KK_{a\bbar} dz^a d\zbar^\bbar + e^{-3 A} \delta_{ij} dy^i dy^j~.}
Given \Hseven, we have that
\eqn\Gthreeexp{G_{3}  = G_{ia\bbar} dy^i dz^a d\zbar^\bbar + \frac{1}{2} G_{ab\cbar} dz^a dz^b d\zbar^\cbar + \frac{1}{2} G_{\abar \bbar c} d\zbar^\abar d\zbar^\bbar dz^c~, \qquad {\rm with}}
\eqn\Gthreecomp{ \eqalign{ &G_{ia\bbar}  = i e^{-3A} \d_i (e^{3A} \KK_{a\bbar})~, \qquad G_{ab\cbar} = 2i e^{-3A} (\d_{[a} e^{3A}) \KK_{b] \cbar}~, \cr
& G_{\abar\bbar c} = -2i e^{-3A} ( {\bar \d}_{[\abar} e^{3A} ) \KK_{|c| \bbar]}~,}}
where we have used the fact that $\KK_{a\bbar} = \d_a {\bar \d}_\bbar \KK$.  Therefore the possible nonzero components of $H_3$ are $H_{ija}, H_{ij\abar}, H_{ia\bbar}$.  We have
\eqn\Hija{ \eqalign{ H_{ija} & = \frac{1}{3!} {\epsilon}_{ijamnp} {g}^{mq} {g}^{nr} {g}^{ps} G_{qrs} \cr
& = \frac{i}{2} e^{-6 A} \varepsilon_{ij}  \epsilon_{ab} \epsilon_{\abar \bbar} g^{b \cbar} g^{\abar d} g^{\bbar e} \left[ (\d_d e^{3A}) g_{e \cbar} - (\d_e e^{3A}) g_{d \cbar} \right] \cr
&= -i \d_a \epsilon_{ij}~. }}
Taking the conjugate gives $H_{ij \abar} = i {\bar \d}_\abar \epsilon_{ij}$.  Finally,
\eqn\Hiabbar{ \eqalign{ H_{i a \bbar} & = \frac{1}{3!} \epsilon_{ia\bbar mnp} {g}^{mq} {g}^{nr} {g}^{ps} G_{qrs} \cr
& = i e^{-3A} \epsilon_{ij} g^{jk} \epsilon_{ac} \epsilon_{\bbar \dbar}  g^{c \ebar} g^{\dbar f} [ (\d_k e^{3A}) g_{f\ebar} + e^{3A} \d_k g_{f \ebar} ]~. }}
Contracting on all available indices, the first term boils down to $i e^{-3A} \epsilon_{i}^{\phantom{i}j} (\d_j e^{3A}) g_{a\bbar}$.  For the second term we have that $\epsilon_{ac} \epsilon_{\bbar \dbar}  g^{c \ebar} g^{\dbar f}  \d_j g_{f \ebar} = \epsilon_{\bbar \dbar} (\d_j g^{\dbar f} ) \epsilon_{f a} \equiv (M_j)_{\bbar a}$, where we've defined a set of $2\times 2$ matrices $M_j$ given by
\eqn\Miab{ \eqalign{ (M_j)_{\bbar a} & = g \left[ \left( \matrix{ 0 &1 \cr -1 & 0} \right) \d_j \left[ g^{-1} \left( \matrix{ g_{2 {\bar 2}} & - g_{1 {\bar 2}} \cr - g_{2 {\bar 1}} & g_{1 {\bar 1}} } \right) \right] \left( \matrix{ 0 & 1 \cr -1 & 0} \right) \right]_{\bbar a} \cr
& = - g (\d_j g^{-1} ) \left( \matrix{ g_{1 {\bar 1}} & g_{2 {\bar 1}} \cr g_{1 {\bar 2}} & g_{2 {\bar 2}} } \right)_{\bbar a} - \d_j \left( \matrix{ g_{1 {\bar 1}} & g_{2 {\bar 1}} \cr g_{1 {\bar 2}} & g_{2 {\bar 2}} } \right)_{\bbar a} \cr
& = -g (\d_j g^{-1}) g_{a \bbar} - \d_j g_{a \bbar}~. }}
where $g={\rm det} g_{a \bar b}$. Collecting results, we find
\eqn\Hiabtwo{ H_{ia \bbar} = i \epsilon_{i}^{\phantom{i}j} \left[ e^{-3A} (\d_j e^{3A}) g_{a\bbar} - g (\d_j g^{-1}) g_{a\bbar} - \d_j g_{a \bbar} \right]~.}
Now, from \gAK\ we have $g = \frac{1}{4} e^{-3A}$.  Using this fact the first two terms in \Hiabtwo\ cancel, and we are left with
\eqn\Hiabthree{ H_{ia\bbar} = - i \epsilon_{i}^{\phantom{i} j} \d_j g_{a\bbar} = -i \epsilon_{i}^{\phantom{i}j} \d_j \d_a {\bar \d}_\bbar \KK~.}
Thus the three-form takes the simple form
\eqn\Hthree{ H_3 = - \frac{i}{2} \left( \d_a \epsilon_{ij} dz^a dy^i dy^j - {\bar \d}_\abar \epsilon_{ij} d\zbar^\abar dy^i dy^j \right) - i \epsilon_{i}^{\phantom{i} j} \d_j \d_a \delbar_\bbar \KK dz^a d\zbar^\bbar dy^i~.}
As a check, we compute $d H_3$ and find
\eqn\dHthree{ \eqalign{ dH_3 =&~  \frac{i}{2} \d_a \delbar_\bbar \left( 2 e^{-3 A} + \Delta_{\bf y} \KK \right) \varepsilon_{ij} dz^a d\zbar^\bbar dy^i dy^j \cr
=&~ - \frac{i }{2}Q_0  \d_a \lambda \delbar_\bbar {\bar\lambda} \ \varepsilon_{ij} \delta^{(2)}({\bf y} - {\bf y}_0) \delta^{(2)}(\lambda - \lambda_0) dz^a d\zbar^\bbar dy^i dy^j~,}}
where \fluxsource\ has been used.  This is precisely what one expects; the $NS5$-brane worldvolume is a magnetic source for $H_3$.

Next we determine a trivialisation for the field strength $H_3$ such that $dB_{2} = H_3$. As there are magnetic sources for $H_3$, viz $dH_3 \ne 0$, the field strength is only trivialisable away from the brane locus. Given that restriction, we ansatz that $B_2$ takes the form
\eqn\Bansatz{ B_{ia} = i \epsilon_{i}^{\phantom{i}j} \d_j v_a = i \varepsilon_{ik} \delta^{kj} \d_j v_a~, \quad ~ B_{i\abar} = -i \epsilon_{i}^{\phantom{i}j} \d_j \vbar_{\abar}~,}
with all other components vanishing and the vector $v_a$ is to be determined.  Differentiating, 
\eqn\dBtwoija{\eqalign{& (dB_2)_{ija} = - i \left( \epsilon_{i}^{\phantom{i} k} \d_j \d_k v_a - \epsilon_{j}^{\phantom{j} k} \d_i \d_k v_a \right) =  -i( \Delta_{\bf y} v_a) \varepsilon_{ij}~, \cr
& (dB_{2})_{ia\bbar} = i \epsilon_{i}^{\phantom{i}j} \d_j (\d_a \vbar_\bbar + {\bar \d}_\bbar v_a)~.}}
Comparing $(dB_2)_{ia\bbar}$ with \Hiabthree, we have
\eqn\va{ (H_3)_{ia\bbar} - (dB_2)_{ia\bbar} = -i \epsilon_{i}^{\phantom{i}j} \d_j \left( \d_a \delbar_\bbar \KK + \d_a \vbar_{\bbar} + \delbar_\bbar v_a  \right) ~.}
Any $v_a$ of the form
\eqn\vafix{ v_a = - \frac{1}{2} \d_a \KK + \frac{1}{2} {\tilde f}({\bf y}) \d_a h(z)~,}
with ${\tilde f}$ an undetermined function of ${\bf y}$ and $h$ a holomorphic function, will cause the right hand side of \va\ to vanish.  Furthermore, with such a $v_a$ the unwanted components $(dB_2)_{iab}, (dB_2)_{i\abar\bbar}$ will vanish.  Plugging this into $(dB_2)_{ija}$ and comparing with \Hija\ gives
\eqn\Hijacompare{  \eqalign{ (H_3)_{ija} - (dB_2)_{ija} =&~  - \frac{i}{2} \d_a \left( 2 e^{-3A} + \Delta_{\bf y} \KK -  h \Delta_{\bf y} {\tilde f} \right) \varepsilon_{ij} \cr
=&~ \frac{i}{2} \left( \frac{Q_0}{2\pi} \delta^{(2)}({\bf y} - {\bf y}_0) \d_a \log{|\lambda - \lambda_0|^2} +  \d_a h \Delta_{\bf y} {\tilde f} \right)~.}  }
The right hand side can be made to vanish by choosing
\eqn\hftilde{ h(z^a) = \log{(\lambda(z^a) - \lambda_0)}~, \qquad \Delta_{\bf y} {\tilde f} = - \frac{Q_0}{2\pi} \delta^{(2)}({\bf y} - {\bf y}_0)~,}
or in the smeared case,
\eqn\ftildesmeared{ \d_{y}^2 {\tilde f} = - \frac{{\tilde Q}_0}{2\pi} \delta(y-y_0)~.}

To summarise the $B$-field is given by
\eqn\Bresult{ B_2 = \frac{i}{2} \epsilon_{i}^{\phantom{i}j} \left(\KK_{ja} dz^a - \KK_{j\abar} d\zbar^\abar \right) dy^i - \frac{i}{2}\left( \star_{\bf y} d_{\bf y} {\tilde f}\,\right) \wedge d \log{\left(\frac{ \lambda - \lambda_0}{{\bar\lambda} - {\bar\lambda}_0} \right)}~,}
with ${\tilde f}$ obeying \hftilde. The second term plays the role of a singular gauge transformation:  it may be written as $d ({\rm Im} \log{(\lambda - \lambda_0)} \star_{\bf y} d_{\bf y} {\tilde f} )$ everywhere except for the surface ${\bf y} = {\bf y}_0$, on which it is no longer exact. The presence of this term is necessary in order for $dB_2 = H_3$ in the plane ${\bf y}={\bf y_0}$, but away from the brane locus $\lambda = \lambda_0$. On the brane locus it is not possible to write $H_3=dB_2$ (as the log is not defined); this is expected, as the branes act as magnetic sources for $H_3$. In Appendix C we will show how these subtleties manifest themselves in the very concrete example of parallel $NS5$-branes and Taub-NUT.

In the text it is convenient to have a more compact expression for the components of $B_2$.  We define
\eqn\Kreg{ \KK_{ia}^{\rm reg.} \equiv \KK_{ia} - \d_i {\tilde f} \d_a \log{(\lambda - \lambda_0)}~,}
and $\KK_{i\abar}^{\rm reg.}$ similarly, so that
\eqn\Bresuseful{ B_2 = \frac{i}{2} \left( \KK_{ja}^{\rm reg.} dz^a - \KK_{j\abar}^{\rm reg.} d\zbar^\abar \right) dy^i~.}
%

\appendix{B}{Supersymmetry of \gcnice\ and comparison with the conifold}

\subsec{Supersymmetry of \gcnice}

The ten-dimensional spacetime of interest is a direct product of flat $\IR^{1,3}$ and the six-dimensional space with metric $d{\tilde s}_{6}^2$, \gcnice, and therefore it is sufficient to analyze the six-dimensional supersymmetry constraints.  Preserved supersymmetries are generated by nontrivial solutions, $\chi$, to the vanishing of the gravitino variation,
\eqn\gravvar{ \delta \Psi_m = \left( \d_m + \frac{1}{4} \omega_{\underline{np},m} \Sigma^{\underline{np}} \right) \chi = 0~.}
We denote the coordinates collectively as $y^m = (y,{\tilde x}^7,v,w,\vbar,\wbar)$.  Underlined indices $\underline{m},\underline{n}$ run over corresponding tangent space directions.  We have $\Sigma^{\underline{mn}} = \Sigma^{[\underline{m}} \Sigma^{\underline{n}]}$, where the $\Sigma^{\underline{m}}$ furnish a representation of the Clifford algebra in $d = 6$ dimensions.

We take the local frame of vielbeins, ${\bf e}^{\underline{m}} = {\bf e}^{\underline{m}}_{\phantom{\underline{m}} n} dy^n$, to be
\eqn\localframe{ \eqalign{ & {\bf e}^{\underline{y}} = e^{-3A/2} dy~, \qquad  {\bf e}^{\underline{7}} = e^{3A/2} \left[ d{\tilde x}^7 - \frac{i}{2} \left( \KK_{ya}^{\rm reg.} dz^a - \KK_{y\abar}^{\rm reg.} d\zbar^{\abar} \right) \right]~, \cr
& {\bf e}^{\underline{v}} = \frac{\sqrt{2}}{\sqrt{\KK_{v\vbar}}} \left( \KK_{\vbar v} dv + \KK_{\vbar w} dw \right)~,  \cr
& {\bf e}^{\underline{w}} =  \frac{\sqrt{2}}{\sqrt{\KK_{v\vbar}}} \sqrt{ \KK_{v\vbar} \KK_{w\wbar} - \KK_{v\wbar} \KK_{w\vbar} } \ dw = \frac{1}{\sqrt{2\KK_{v\vbar}}} e^{-3A/2} dw~,}}
with ${\bf e}^{\underline{\vbar}},{\bf e}^{\underline{\wbar}}$ given by conjugation.  One may verify that these satisfy ${\bf e}^{\underline{m}}_{\phantom{\underline{m}} m} {\bf e}^{\underline{n}}_{\phantom{\underline{n}} n} \eta_{\underline{mn}} = g_{mn}$, where $\eta_{\underline{mn}}$ is the flat tangent space metric and our conventions are $\eta_{\underline{v\vbar}} = \half$, $\eta_{\underline{vv}} = \eta_{\underline{\vbar\vbar}} = 0$, and similarly for $\underline{w}$.   The computation of the spin connection $\omega$ is completely straightforward, though tedious, and the details will not be recorded here.  After repeated use of \gAK\ and the equation of motion, \Mongenice, which appears in the form
\eqn\eomder{ \d_y \KK_{ya}^{\rm reg.} + 2 \d_a (e^{-3A}) = 0~, \qquad \d_y \KK_{y\abar}^{\rm reg.} + 2 \delbar_\abar (e^{-3A}) = 0~,}
we find that
\eqn\label{ \omega_{\underline{np},m} \Sigma^{\underline{np}} \chi = 0~, \quad \forall ~m~,}
provided $\chi$ satisfies the following two projection conditions:
\eqn\projectors{ \half \left( {\bf 1} + \frac{i}{2} \Sigma^{\underline{y7}} \Sigma^{\underline{v\vbar}} \right) \chi = \chi =  \half \left( {\bf 1} + \frac{i}{2} \Sigma^{\underline{y7}} \Sigma^{\underline{w\wbar}} \right) \chi~.}
Thus any {\it constant} spinor $\chi$, satisfying \projectors, will solve \gravvar.  Since each projection condition removes half of the degrees of freedom, the geometry is $1/4$-BPS.

It will be convenient to summarise results in a specific basis.  Let us take
\eqn\gammabasis{ \Sigma^{\underline{y},\underline{7}} = \sigma^{1,2} \otimes {\bf 1}_2 \otimes {\bf 1}_2~, \quad \Sigma^{\underline{v},\underline{\vbar}} = \sigma^3 \otimes \sigma^{+,-} \otimes {\bf 1}_2~, \quad \Sigma^{\underline{w},\underline{\wbar}} = \sigma^3 \otimes \sigma^3 \otimes \sigma^{+,-}~,}
with $\sigma^i$ the standard Pauli matrices and
\eqn\sigmapm{ \sigma^{\pm} = \sigma^1 \pm i \sigma^2~.}
Then \projectors\ reads
\eqn\projectorsb{ \half \left( {\bf 1}_8 - \sigma^3 \otimes \sigma^3 \otimes {\bf 1}_2 \right) \chi = \chi =  \half \left( {\bf 1}_8 - \sigma^3 \otimes {\bf 1}_2 \otimes \sigma^3 \right) \chi~,}
and the general solution may be written as a linear combination of $\chi_{\pm}$, where
\eqn\chipm{ \chi_+ = | \uparrow \uparrow \uparrow \rangle~, \qquad \chi_- = | \downarrow \downarrow \downarrow \rangle~.}
Here we use a standard quantum mechanics notation where $\sigma^3 | \uparrow \rangle = | \uparrow \rangle$, $\sigma^3 | \downarrow \rangle = - | \downarrow \rangle$, and $\sigma^+ | \uparrow \rangle = \sigma^- | \downarrow \rangle = 0$.  

\subsec{The conifold limit}

We would like to show that the above results, valid for any $\KK$ satisfying \Mongenice, are consistent with known results in the special case of the conifold.  Let us first recall what the known results are.  We work with coordinates ${y'}^{m'} = (\rho,\psi,\theta_1,\phi_2,\theta_2,\phi_2)$, in terms of which the metric is given by \conemet.  Taking a basis of vielbeins ${\bf e'}^{\underline{m}'} = {\bf e'}^{\underline{m}'}_{\phantom{\underline{m}'} n'} d{y'}^{n'}$, given by
\eqn\conbeins{ \eqalign{ & {\bf e'}^{\underline{\rho}} = d\rho~, \qquad {\bf e'}^{\underline{\psi}} = \frac{\rho}{3} \left[ d\psi + \cos{\theta_1} d\phi_1 + \cos{\theta_2} d\phi_2 \right]~, \cr
& {\bf e'}^{\underline{\theta_i}} = \frac{\rho}{\sqrt{6}} d\theta_i~, \qquad {\bf e'}^{\underline{\phi_i}} = \frac{\rho}{\sqrt{6}} \sin{\theta_i} d\phi_i~,}}
one finds that the Killing spinors are linear combinations of
\eqn\chiconpmone{ \chi_{\pm}' = e^{\pm i \psi/2} \chi_{0\pm}'~,}
where $\chi_{0\pm}'$ are constant spinors satisfying
\eqn\conproj{ \half \left( {\bf 1} + \Sigma^{\underline{\rho\psi}} \Sigma^{\underline{\theta_1 \phi_1}} \right) \chi_{0\pm}' = \chi_{0\pm}' = \half \left( {\bf 1} + \Sigma^{\underline{\rho\psi}} \Sigma^{\underline{\theta_2 \phi_2}} \right) \chi_{0\pm}'~,}
and with $-i \Sigma^{\underline{\rho\psi}} \chi_{0\pm}' = \pm \chi_{0\pm}'$.  In terms of the basis
\eqn\conbasis{ \Sigma^{\underline{\rho},\underline{\psi}} = \sigma^{1,2} \otimes {\bf 1}_2 \otimes {\bf 1}_2~, \quad \Sigma^{\underline{\theta_1}, \underline{\phi_1}} = \sigma^3 \otimes \sigma^{1,2} \otimes {\bf 1}_2~, \quad \Sigma^{\underline{\theta_2},\underline{\phi_2}} = \sigma^3 \otimes \sigma^3 \otimes \sigma^{1,2}~,}
we have
\eqn\chiconpm{ \chi_{+}' = e^{i\psi/2} | \uparrow \downarrow \downarrow \rangle~, \qquad \chi_{-}' = e^{-i\psi/2} | \downarrow \uparrow \uparrow \rangle~.}

Our goal is to demonstrate the equivalence of \chipm\ and \chiconpm.  To see how this is possible, we must recall that in general, two sets of vielbeins ${\bf e}^{\underline{m}}_{\phantom{\underline{m}} n}$ and ${\bf e'}^{\underline{m}'}_{\phantom{\underline{m}'} n'}$ are related by both a coordinate transformation and local frame rotation $\Lambda \in SO(6)$:
\eqn\vieltrans{ {\bf e'}^{\underline{m}'}_{\phantom{\underline{m}'} n'} = \Lambda^{\underline{m}'}_{\phantom{\underline{m}'} \underline{m}}  {\bf e}^{\underline{m}}_{\phantom{\underline{m}} n} \frac{\d y^n}{\d  {y'}^{n'}}~.}
The gravitino variation \gravvar\ must transform covariantly under both general coordinate transformations and local frame rotations.  While the spinor $\chi$ transforms trivially (as a scalar) under coordinate transformations, the action of the frame rotation is nontrivial:
\eqn\chitrans{ \chi' = \Lambda_{\half} \chi~,}
where $\Lambda_\half$ is constructed by exponentiating the generators in the spinor representation with the same coefficients used in constructing $\Lambda$ via exponentiation of vector representation generators.

Determining $\Lambda$ is straightforward, since we have explicit expressions for the other three matrices, ${\bf e}'$, ${\bf e}$, $\d y/\d y'$ appearing in \vieltrans.  The coordinate transformation, for example, is (working in the upper patch)
\eqn\coordtransap{ \eqalign{ & y = \frac{\rho^2}{6 {\tilde R}_7} (\cos{\theta_1} + \cos{\theta_2})~, \qquad  {\tilde x}^7 = \frac{{\tilde R}_7}{2} (\psi + \phi_1 + \phi_2)~, \cr
& v = - \frac{2 \rho^{3/2}}{3 {\tilde R}_{7}^{1/2}} \cos{\frac{\theta_1}{2}} \sin{\frac{\theta_2}{2}} e^{\frac{i}{2} (\psi + \phi_1 - \phi_2)}~, \cr
& w = \frac{2 \rho^{3/2}}{3 {\tilde R}_{7}^{1/2}} \sin{\frac{\theta_1}{2}} \cos{\frac{\theta_2}{2}} e^{\frac{i}{2} (\psi - \phi_1 + \phi_2)} ~.}}
One technical point worth mentioning is that in order for $\Lambda$ to be an element of $SO(6)$, such that $\Lambda^T \Lambda = {\bf 1}$, it is necessary to work in a real basis of vielbeins, ${\bf e}^{\underline{v_1},\underline{v_2}}, {\bf e}^{\underline{w_1},\underline{w_2}}$, related to the holomorphic basis by ${\bf e}^{\underline{v}} = {\bf e}^{\underline{v_1}} + i {\bf e}^{\underline{v_2}}$, \etc.  In a slight abuse of notation we have used the index $\underline{m}$ in \vieltrans\ to run over the values $(\underline{y},\underline{7},\underline{v_1},\underline{v_2},\underline{w_1},\underline{w_2})$.

We find that $\Lambda$ takes the form
\eqn\Lambdafactor{ \Lambda = - {\tilde \Lambda} \cdot R_{\phi}~,}
where $R_\phi$ consists of rotations in the $v$- and $w$-planes,
\eqn\Rphi{ R_\phi = e^{\frac{i}{2} (\psi + \phi_1 - \phi_2) J_{\underline{w_1 w_2}}} \cdot e^{\frac{i}{2} (\psi - \phi_1 + \phi_2) J_{\underline{v_1 v_2}}}~.}
If we order the coordinate axes according to $(\underline{y},\underline{v_1},\underline{w_1},\underline{7},\underline{v_2},\underline{w_2})$ and $(\underline{\rho},\underline{\theta_1},\underline{\theta_2},\underline{\psi},\underline{\phi_1},\underline{\phi_2})$, then ${\tilde \Lambda}$ takes a block diagonal form, consisting of two $SO(3)$ rotations:
\eqn\Lambdatilde{ {\tilde \Lambda} = e^{i\pi J_{\underline{v_2w_2}}} \cdot e^{i\beta_3 (J_{\underline{v_2 w_2}} + J_{\underline{v_1 w_1}})} \cdot e^{i\beta_2 (J_{\underline{7 v_2}} + J_{\underline{y v_1}})} \cdot e^{i\beta_1 (J_{\underline{v_2 w_2}} + J_{\underline{v_1 w_1}})}~.}
In these expressions $J_{\underline{mn}}$ are the rotation generators in the vector representation.  The $\beta_i$ are $SO(3)$ Euler angles; they are rather nontrivial functions of $(\theta_1,\theta_2)$.  Their explicit form can be given but will not be needed in the following.  The important point is that the two $SO(3)$ rotations are identical except for a relative shift of $\beta_3 \to \beta_3 + \pi$ on the third Euler angle.  $\Lambda_{\half}$ is constructed in an identical fashion, but with $J_{\underline{mn}} \to \JJ_{\underline{mn}}$, where
\eqn\spinorgen{ \JJ_{\underline{mn}} = - \frac{i}{4} \left[ \Sigma_{\underline{m}} , \Sigma_{\underline{n}} \right]~.}
In particular, we define $(R_\phi)_\half$ and ${\tilde \Lambda}_\half$ such that $\Lambda_\half = - (R_\phi)_\half \cdot {\tilde \Lambda}_\half$.

Now consider the action of $\Lambda_\half$ on $\chi_{\pm}$, \chipm.  First we have that
\eqn\Rphionchi{ (R_\phi)_\half \chi_\pm = \left[ {\bf 1}_2 \otimes e^{\frac{i}{4}(\psi +\phi_1 - \phi_2) \sigma^3} \otimes e^{\frac{i}{4} (\psi - \phi_1 + \phi_2) \sigma^3} \right] \chi_{\pm} = e^{\pm i\psi/2} \chi_{\pm}~.}
Meanwhile, observe that
\eqn\Jtrivial{ \eqalign{ &\JJ_{\underline{v_2 w_2}} + \JJ_{\underline{v_1 w_1}} = -\frac{i}{4} {\bf 1}_2 \otimes \left[ \sigma^- \otimes \sigma^+ - \sigma^+ \otimes \sigma^- \right]~, \cr
& \JJ_{\underline{7 v_2}} + \JJ_{\underline{y v_1}} = - \frac{i}{4} \left[ \sigma^- \otimes \sigma^+ - \sigma^+ \otimes \sigma^- \right] \otimes {\bf 1}_2~.}}
Since these expressions annihilate $\chi_{\pm}$, the analogous $\beta$-dependent pieces of \Lambdatilde\ in ${\tilde \Lambda}_\half$ act trivially.  Finally, since $e^{i\pi \JJ_{\underline{v_2 w_2}}} = i \left[ {\bf 1}_2 \otimes \sigma^1 \otimes \sigma^2 \right]$, we have
\eqn\Lambdaact{ \Lambda_\half \chi_+ = e^{i\psi/2} | \uparrow \downarrow \downarrow \rangle = \chi_{+}'~, \qquad \Lambda_{\half} \chi_- = -e^{-i\psi/2} |\downarrow \uparrow \uparrow \rangle = - \chi_{-}'~.}
This demonstrates that the Killing spinors of \gcnice\ are consistent with the standard results for the Killing spinors of the conifold, \chiconpm.

\appendix{C}{Taub-NUT limit}

A special case of the $D5$-brane system \Dfive\ is when we have a single stack of parallel $D5$-branes. $S$-duality maps this to a stack of parallel $NS5$-branes; then $T$-duality on the transverse circle brings one to multi-centered Taub-NUT.  Hence, multi-centered Taub-NUT should be a special case of the general metric \gcnice.  Let's start with the simplest case of (one-centered) Taub-NUT,  and see how this is embedded in \gcnice.

We label the holomorphic coordinates $z^1 = v$, $z^2 = w$, and we will take $w$ to be the direction along which the $NS5$-branes are extended.  One expects a completely flat geometry in $\IC_{w}$, and no mixing between $y,w$ or $v,w$.  Therefore we make the ansatz
\eqn\TNansatzone{ \KK_{w\wbar} = \frac{1}{2}~, \qquad \KK_{v \wbar} = \KK_{w \vbar} = 0~, \qquad \KK_{y w} = \KK_{y \wbar} = 0~,}
in which case \gcnice\ becomes
\eqn\TNsimpone{ d{\tilde s}_{6}^2 = e^{-3A} dy^2 + 2 \KK_{v \vbar} dv d\vbar + e^{3A} \left( d{\tilde x}^7 - \frac{i}{2} (\KK_{yv}^{\rm reg.} dv - \KK_{y\vbar}^{\rm reg.} d\vbar ) \right)^2 + dw d\wbar~.}

Recall the form of the Taub-NUT metric is
\eqn\TN{ \eqalign{ ds_{TN}^2 =&~ V ( dy^2 + dv d\vbar ) + \frac{1}{V} ( dx^7 + \omega_{v} dv + \omega_{\vbar} d\vbar )^2~, \qquad {\rm where} \cr
\star d\omega &=~ dV~, \qquad {\rm with} \quad V(y,v,\vbar) = 1 + \frac{{\tilde R}_7}{2 \sqrt{ y^2 + |v|^2}}~.}}
The hodge dual is taken with respect to the flat $\IR^{3}$ metric, $dy^2 + dv d\vbar$, and we have made a gauge choice for the one-form $\omega$ so that it has no legs along $y$.  ${\tilde R}_7$ is the radius of the ${\tilde x}^7$ circle fiber at infinity.  Let us introduce spherical coordinates,
\eqn\sphericalcoord{ (v,y) = (r \sin{\theta} e^{i\phi}, r \cos{\theta} )~.}
For fixed $r$, $({\tilde x}^7,\omega)$ gives a nontrivial circle fibration (the Hopf fibration) over $S^2$.  It can be described using two standard coordinate patches:
\eqn\omegapm{ \eqalign{& \omega^+ = - \frac{{\tilde R}_7}{2} (1 - \cos{\theta}) d\phi~, \qquad  \theta \in [0,\pi)~, \cr
& \omega^- = \frac{{\tilde R}_7}{2} (1 + \cos{\theta}) d\phi~, \qquad  \theta \in (0,\pi]~.}}
On the overlap, the transition $\omega^- = \omega^+ + {\tilde R} d\phi$ is compensated by a (periodicity-preserving) coordinate transformation
\eqn\Taubfibre{ {\tilde x}^{7(-)} = {\tilde x}^{7(+)} - {\tilde R}_7 \phi~.}
The circle fibre is well defined everywhere except at $r = 0$, where neither $\omega^{\pm}$ make sense.  Fortunately the fibre shrinks to zero there.  The two-sphere shrinks at the same rate so that we get the metric on a round $S^3$ as $r \to 0$ and the total space is smooth, with global topology $\IR^4$.  In the dual picture, $r=0$ corresponds to the location of the $NS5$-brane.

In order to put \TNsimpone\ in the form of \TN\ (plus a flat $\IC_w$), we  require
\eqn\TNansatztwo{ e^{-3A} = 2 \KK_{v \vbar} = V~, \qquad - \frac{i}{2} (\KK_{yv}^{\rm reg.})^\pm = \omega_{v}^\pm \quad  \frac{i}{2} (\KK_{y\vbar}^{\rm reg.})^\pm = \omega_{\vbar}^\pm~.}
Note that it is consistent to have $\KK_{yv}^{\rm reg.},\KK_{y\vbar}^{\rm reg.}$ patch dependent, while the other $\KK_{mn}$ are patch independent, provided that $\KK^{\pm}$ are related by a gauge transformation of the form \KKahlertrans.  Furthermore, since the warp factor already takes the canonical near-brane form, we may in this case trivially identify the near-brane coordinates,
\eqn\nearbraneTN{ (\lambda,\eta) = (v,w)~.}

The relations \TNansatztwo, \TNansatzone\ must be consistent with both $\star d\omega = dV$, and the Monge-Ampere equation, \Mongenice.  Let us first assume that \Mongenice\ is satisfied and check the relation between $\omega$ and $V$.  We have
\eqn\dVwithK{ dV = \left(  \d_v e^{-3A} dv +\d_\vbar e^{-3A} d\vbar + \d_y e^{-3A} dy \right)~,}
while
\eqn\stardomegawithK{ \eqalign{ \star d \omega =&~ \epsilon_{vmn} \eta^{mp} \eta^{nq} \d_p \omega_q dv + \epsilon_{\vbar mn} \eta^{mp} \eta^{nq}  \d_p \omega_q d\vbar+ \epsilon_{y mn} \eta^{mp} \eta^{nq} \d_p \omega_q dy \cr
=&~ i ( \d_v \omega_y - \d_y \omega_v )dv + i( \d_y \omega_\vbar - \d_\vbar \omega_y) d\vbar + 2i (\d_\vbar \omega_v - \d_v \omega_\vbar ) dy \cr
=&~ - \frac{1}{2} \KK_{yyv}^{\rm reg.} dv - \frac{1}{2} \KK_{yy\vbar}^{\rm reg.} d\vbar +2 \KK_{yv\vbar}^{\rm reg.} dy~.}}
Subtracting the two results and using \Kreg\ gives
\eqn\dVstarcheck{ \eqalign{ dV - \star d\omega =&~ \frac{1}{2} \d_v \left( 2 e^{-3A} + \KK_{yy} - \d_{y}^2 {\tilde f} \log{(v)} \right) dv + \cr
& + \frac{1}{2} \delbar_\vbar \left( 2 e^{-3A} + \KK_{yy} -  \d_{y}^2 {\tilde f} \log{(\vbar)} \right) d\vbar + \cr
& +  \d_y \left( e^{-3A} - 2 \KK_{v\vbar} \right) dy~.}}
The last line trivially vanishes assuming \TNansatztwo.  We see that the first two lines vanish as well, using \Mongenice\ and \ftildesmeared.  Thus $dV = \star d\omega$ is satisfied, provided $\KK$ satisfies Monge-Ampere.  Notice the importance of having $\omega_{v} \sim \KK_{yv}^{\rm reg.}$ and not $\KK_{yv}$.  If we had used a $B_2$ to construct the T-duality that did not satisfy $dB_2 = H_3$ in the $y =0$ plane, \TNsimpone\ would not have reproduced the proper circle fibre for Taub-NUT in the $y=0$ plane.

Now let's consider the Monge-Ampere equation.  First, away from $y=0$ we have
\eqn\Kyyynotzero{ \KK_{yy} = -2 e^{-3A} = -2 V~, \qquad (y \neq 0)~.}
This, combined with \TNansatzone, \TNansatztwo\ implies that the source-free equation holds as an algebraic equation for unrelated functions $\KK_{mn}$.  By direct calculation, one verifies that either of the functions,
\eqn\KpmTN{ \KK^{\pm} = \frac{1}{2}(v\vbar + w\wbar - 2 y^2) + {\tilde R}_7 \left\{ \sqrt{y^2 + |v|^2} \mp y \log{ \left[ \pm y + \sqrt{y^2 + |v|^2} \right]} \right\}~,}
satisfies the source-free equation.  In fact, any linear combination of the form $\alpha \KK^+ + \beta \KK^-$, with $\alpha + \beta = 1$ satisfies the equation since all of these solutions are related by gauge transformations of the form \KKahlertrans.  For $v \neq 0$ either of these expressions is suitable, but for $v = 0, y >0$ we must use $\KK^+$, while for $v = 0, y <0$ we must use $\KK^-$.  This defines a solution to \Mongenice\ everywhere away from $y = 0$.

There is a unique way to extend the solution to $y=0$ such that it provides the right singularity structure:
\eqn\KTN{ \KK_{TN} =  \frac{1}{2}(v\vbar + w\wbar - 2 y^2) + {\tilde R}_7 \left\{ \sqrt{y^2 + |v|^2} - |y| \log{ \left[ |y| + \sqrt{y^2 + |v|^2} \right]} \right\}~.}
We only require $\KK_{TN}$ to take this form as $y \to 0$; away from $y=0$ it amounts to a convenient gauge choice that allows for a global definition of $\KK$.  As we know from \Kexpansion--\dLambdaKtwo, this expression satisfies the sourced Monge-Ampere equation with the expected charge of ${\tilde Q} = 2\pi {\tilde R}_7$.  Computing the mixed derivatives, $\KK_{yv}$ and $\KK_{y\vbar}$, leads to
\eqn\singularfibre{ \eqalign{ - \frac{i}{2} ( \KK_{y v} dv - \KK_{y\vbar} d\vbar ) =&~ -\frac{i {\tilde R}_7}{4} \left( \frac{y}{\sqrt{y^2 + |v|^2}} - \sgn(y) \right) d \log{(v/\vbar)} \cr
=&~ \frac{{\tilde R}_7}{2} ( \cos{\theta} - \sgn(y)) d\phi~.}}
By taking
\eqn\tildefpm{ {\tilde f}^{\pm} = \frac{{\tilde R}_7}{2} \left(- |y| \pm y \right)~,}
which satisfies \ftildesmeared\ with ${\tilde Q}_0 = 2\pi {\tilde R}_7$, we get the one-forms \omegapm:
\eqn\regularfibre{ \omega^{\pm} = - \frac{i}{2} \left( ( \KK_{y v}^{\rm reg.})^{\pm} dv - (\KK_{y\vbar}^{\rm reg.})^{\pm} d\vbar \right) = \frac{{\tilde R}_7}{2} \left( \cos{\theta} \mp 1 \right) d\phi~.}

This completes our demonstration that \gcnice\ describes Taub-NUT as a special case.  Using \KTN, one may also verify that \NSfive, \NSHthree\ describes the H-monopole solution \refs{\BanksRJ,\CallanAT}, for one $NS5$-brane smeared on a transverse circle.

Finally, since the Monge-Ampere equation reduced to a linear PDE due to the large amount of symmetry present, we can use the superposition principle.  The generalisation of \KTN\ to multi-centered Taub-NUT is trivial:
\eqn\multicenter{  \eqalign{ & \KK_{TN} = \frac{1}{2} \left( v\vbar + w\wbar - 2 y^2 \right) + {\tilde R}_7 \sum_{i} \KK^{(i)}~, \qquad {\rm where} \cr
& \KK^{(i)} =\sqrt{(y-y_i)^2 + |v-v_i|^2} - |y-y_i| \log{ \left[ |y-y_i| + \sqrt{(y-y_i)^2 + |v-v_i|^2} \right]}~.}}

\listrefs
\end